\documentclass[superscriptaddress,amsmath,amssymb,aps,reprint]{revtex4-1}

\usepackage{graphicx}
\usepackage{dcolumn}
\usepackage{bm}
\usepackage[hypertexnames=false]{hyperref}
\usepackage{notes2bib}
\usepackage{siunitx}
\usepackage{amsmath}
\usepackage[usenames,dvipsnames]{xcolor}
\usepackage[utf8]{inputenc}

\usepackage{soul}

\usepackage{lineno}
\newcommand*{\Iex}{I_{\mathrm{ex}}}
\newcommand*{\Ic}{I_{\mathrm{c}}}
\newcommand*{\Isw}{I_{\mathrm{sw}}}
\newcommand*{\Ire}{I_{\mathrm{re}}}

\newcommand*{\Gammare}{\Gamma_{\mathrm{re}}}
\newcommand*{\Gammasw}{\Gamma_{\mathrm{sw}}}
\newcommand*{\Gth}{G_{\mathrm{th}}}
\newcommand*{\Cel}{C_{\mathrm{el}}}
\newcommand*{\Lkin}{L_{\mathrm{kin}}}
\newcommand*{\Rbulk}{R_{\mathrm{bulk}}}
\newcommand*{\Rj}{R_{\mathrm{J}}}
\newcommand*{\Isc}{I_{\mathrm{sc}}}
\newcommand*{\Ij}{I_{\mathrm{J}}}
\newcommand*{\Tj}{T_{\mathrm{J}}}
\newcommand*{\ns}{n_{\mathrm{s}}}

\begin{document}

\title{Quasiparticle and superfluid dynamics in Magic-Angle Graphene}

\author{El\'ias Portol\'es}
\email{eliaspo@phys.ethz.ch}
\author{Marta Perego}
\affiliation{Laboratory for Solid State Physics, ETH Zurich,~CH-8093~Zurich, Switzerland}
\author{Pavel A. Volkov}
\email{pavel.volkov@uconn.edu}
\affiliation{Department of Physics, University of Connecticut, Storrs, Connecticut 06269, USA}
\affiliation{Department of Physics, Harvard University, Cambridge, Massachusetts 02138, USA}
\author{Mathilde Toschini}
\author{Yana Kemna}
\author{Alexandra Mestre-Tor\`a}
\author{Giulia Zheng}
\author{Artem O. Denisov}
\author{Folkert K. de Vries}
\author{Peter Rickhaus}
\affiliation{Laboratory for Solid State Physics, ETH Zurich,~CH-8093~Zurich, Switzerland}
\author{Takashi Taniguchi}
\affiliation{International Center for Materials Nanoarchitectonics, National Institute for Materials Science,  1-1 Namiki, Tsukuba 305-0044, Japan}
\author{Kenji Watanabe}
\affiliation{Research Center for Functional Materials, National Institute for Materials Science, 1-1 Namiki, Tsukuba 305-0044, Japan}
\author{J. H. Pixley}
\affiliation{Department of Physics and Astronomy, Center for Materials Theory, Rutgers University, Piscataway, New Jersey 08854, USA}
\affiliation{Center for Computational Quantum Physics, Flatiron Institute, 162 5th Avenue, New York, NY 10010}
\author{Thomas Ihn}
\author{Klaus Ensslin}
\affiliation{Laboratory for Solid State Physics, ETH Zurich,~CH-8093~Zurich, Switzerland}
\affiliation{Quantum Center, ETH Zurich,~CH-8093 Zurich, Switzerland}

\begin{abstract}

Magic-Angle Twisted Bilayer Graphene shows a wide range of correlated phases which are electrostatically tunable.
Despite a growing knowledge of the material \cite{Stepanov2020, Lu2019, Oh2021}, there is yet no consensus on the microscopic mechanisms driving its superconducting phase \cite{Kennes2018, Biao2019, Qin2023}.
In particular, elucidating the symmetry and formation mechanism of the superconducting phase may provide key insights for the understanding of unconventional, strongly coupled and topological superconductivity.
A major obstacle to progress in this direction is that key thermodynamic properties, such as specific heat, electron-phonon coupling and superfluid stiffness, are extremely challenging to measure due to the 2D nature of the material and its relatively low energy scales.
Here, we use a gate-defined, radio frequency-biased, Josephson junction to probe the electronic dynamics of magic-angle twisted bilayer graphene (MATBG).
We reveal both the electronic quasiparticle dynamics, driven by their thermalization through phonon scattering, as well as the condensate dynamics, driven by the inertia of Cooper pairs.
From these properties we recover the evolution of thermalization rates, and the superfluid stiffness across the phase diagram.
Our findings favor an anisotropic or nodal pairing state and allow to estimate the strength of electron-phonon coupling.
These results contribute to understanding the underlying mechanisms of superconductivity in MATBG while establishing an easy-to-implement method for characterizing thermal and superfluid properties of superconducting 2D materials.

\end{abstract}

\maketitle

%Introduction 
The phase diagram of magic-angle twisted bilayer graphene \cite{Cao2018_1, Cao2018_2} (MATBG) has drawn considerable attention due to the presence of correlated insulating, superconducting, and topological phases \cite{Lu2019}.
Despite these  remarkable discoveries, several questions remain open about the nature of the superconducting state and even less is understood about what is the driving mechanism behind it.
The most central and pressing issues include whether the superconducting mechanism of MATBG is electronic or phonon-driven \cite{sarma2019,Biao2019, WuIvar-2018, Kennes2018} and whether its superconducting gap is nodal or not \cite{Oh2021, Kim2022}. 
Furthermore, the electron-phonon coupling has been suggested as the origin of the observed (putatively) universal linear-in-T resistance \cite{sarma2019,polshyn2019large}.
Characterizing the electron-phonon coupling and the anisotropy of the superconducting gap is the first step towards answering these questions. 
However, the 2D nature of MATBG, the small moir\'e Brillouin zone, and the relatively low energy scales make the use of many standard techniques for investigating bulk materials, such as calorimetry, ARPES or neutron scattering, challenging or impossible.

\begin{figure*}[t!]
\includegraphics[width=1\textwidth]{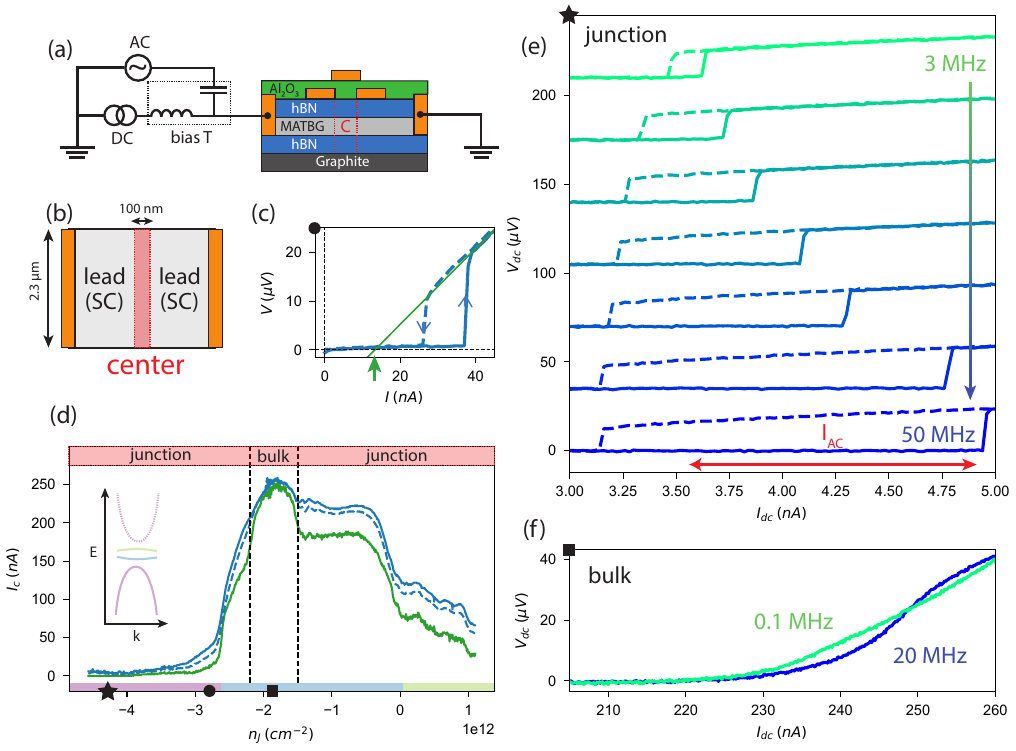}
\caption{(a) Schematics of the device. The device is depicted by a cross section schematics. (b) Top-view simplified schematics with the gold contacts on the side, connected by a stripe of MATBG. The central region, of length $\SI{100}{nm}$, is highlighted. (c) I/V characteristic of the junction at a density of $\SI{-4.3e-12}{cm^{-2}}$. In blue a trace for increasing DC bias is shown, in orange for decreasing bias. The green line is an extrapolation of the resistive part of the characteristic at 0 voltage. (d) Switching (blue), retrapping (blue, dashed) and excess (green) currents as a function of density in the central region. The colors on the x-axis correspond to the filling of the band structure schematics shown in the inset. The upper part indicates whether the IV characteristic shows a junction-like or bulk superconductor-like behavior.
The black star, circle and square indicate, respectively, the densities at which is taken the data shown in (e), (c) and (f). (e) I/V traces of the junction for AC bias of increasing frequency and fixed amplitude (red arrow) as a function of DC bias (horizontal axis).
Solid lines show positive bias directions while dashed ones show negative directions. Curves are offset vertically for readability. (f) I/V traces at bias AC amplitude $\SI{1.4}{nA}$ and frequencies of $\SI{0.1}{MHz}$ and $\SI{20}{MHz}$ when the sample is tuned to all-bulk configuration see panel (d).}
\label{fig:1}
\end{figure*}

%Small paragraph saying how useful the combination of superconducting devices with RF is
Superconducting mesoscopic devices have proven to be a useful characterization tool of the material they are built of \cite{Wollman1993, Lopez2023, Chiodi2008}.
In particular, Josephson junctions (JJs) have been used as a probe of electronic thermalization rates~\cite{Chiodi2008}, and the superfluid density, through characterizing the kinetic inductance in thin-film devices~\cite{Lopez2023}.
In the case of MATBG, superconducting devices have already proven instrumental for probing the charge of the Cooper pairs \cite{deVries2021,Portolés2022}, the long-range coherence of the superconducting condensate \cite{Portolés2022} and its orbital magnetic properties \cite{Díez-Mérida2023}.

%Here, we...
Here, we use a gate-defined Josephson junction (JJ) in MATBG \cite{deVries2021, Rodan-Legrain2021, Díez-Mérida2023} to extract electron-phonon coupling, thermodynamic, and superfluid properties of MATBG across its phase diagram. 
Biasing the junction with a combination of DC and AC currents we probe the dynamics of both the electronic quasiparticles and the superfluid of MATBG.
We show that the measured timescales governing the junction's transition between resistive and superconducting states are directly related to the microscopic properties of the material, such as electronic cooling power due to phonons, specific heat, and superfluid density.
The gate-tunability of the device allows us to probe these quantities across the density-tuned phase diagram of MATBG, for chemical potential both within and outside the flat bands.
These measurements imply that MATBG's electron-phonon coupling is weak, e.g. lower than that of aluminium (a conventional superconductor), and the current bias dependence of the superfluid density is incompatible with isotropic pairing.

%Device description
Our device is a JJ electrostatically defined in MATBG, with a twist angle of $1.06^\circ \pm 0.04^\circ$, also studied in reference \cite{deVries2021} (Fig.~\ref{fig:1}(a)).
The global carrier density $n$, tuned by the back gate, is set to $n = \SI{-1.73e-12}{cm^{-2}}$, at which the bulk has its highest critical current, $\SI{250}{nA}$ (See SI).
Two layers of top gates, separated by a layer of $\mathrm{Al_2O_3}$ tune the local density in the central region, allowing us to fine-tune the details of the junction.

%From the hysteresis of the junction we can probe the dynamics of superfluid and quasiparticles
For each value of electron density in the central region (Fig.~\ref{fig:1}(b)) we analyze the current-voltage (I/V) characteristic.
For densities in the central region close to $\nu=-2$ we observe a gradual onset of resistance above a critical current value, consistent with bulk superconductivity (see also discussion of Fig.~\ref{fig:1}(f) below).
For all other densities, we universally observe a hysteretic I/V trace with two characteristic voltage jumps $\Delta V$, as shown in Fig.~\ref{fig:1}(c).
The two jumps correspond to switching from the superconducting to the resistive state (increasing current bias, blue line) and retrapping back (decreasing current bias, blue dashed).
Together with Shapiro step measurements \cite{deVries2021} this indicates the formation of a weak superconducting link between the left and right parts of the device, where the weak link region can switch between resistive and superconducting states.
From the band structure of MATBG \cite{Cao2018_2} (see inset of Fig.~\ref{fig:1}(d), for a schematic), the weak link region is expected to be metallic except for a narrow range of voltages placing the chemical potential into the gap between the flat and dispersive bands.
Such assessment is consistent with the observation of a positive excess current \cite{Tinkham2004,BTK} in the resistive state of a large portion of the phase diagram (green curve in Fig.~\ref{fig:1}(d)).
In analogy to conventional superconductors \cite{Chiodi2008}, the dynamic response of such metallic weak links should give access to the dynamics of the electronic quasiparticles and the superconducting condensate in MATBG.

%We probe the junction with RF. The junction behaves as expected
We probe the dynamics of our weak links by adding a small AC current component to the DC current flowing through the junction.
Sweeping the frequency across three orders of magnitude (0.1-100 MHz), we focus on the changes in the I/V characteristics, as shown in Fig.~\ref{fig:1}(e). 
At low frequencies, the AC drive brings the two hysteresis branches closer together, which can be understood as follows.
The abrupt character of switching and retrapping with DC bias suggests that the junction will undergo a change whenever the total current $I_{DC}+I_{RF}(t)$ reaches the critical value for switching ($\Isw$) or retrapping ($\Ire$). 
Consequently, one expects the switching to occur prematurely at $\Isw-I_{RF}$, and the retrapping to occur at a higher DC bias, $\Ire+I_{RF}$, reducing the size of the hysteresis loop.

%Switching and retrapping rates of the junction tell us physical properties about MATBG
For increasing frequency, the effect of AC bias gradually disappears (Fig.~\ref{fig:1}(e)), with a different rate for switching and retrapping.
This indicates that both processes, in fact, do not occur instantaneously and are characterized each by a certain rate, which we denote as $\Gammare$ and $\Gammasw$ with precise definitions introduced below.
At highest frequencies, the AC drive effect is absent  (See Fig.~\ref{fig:3}(c) and Supplemental Material), indicating that neither switching nor retrapping processes are fast enough to occur over one AC drive period.
The switching and retrapping rates that can be extracted from Fig.~\ref{fig:1}(e) reflect the physical properties of superconducting MATBG. 
We now turn to their physical interpretation.

%We rule out RCSJ
We can first rule out switching and retrapping driven only by the dynamics of the superconducting phase difference across the junction, exemplified by, e.g.,the  RCSJ model \cite{Tinkham2004}.
In that case, the characteristic frequency is fixed by the Josephson relation to  $2e \Delta V/\hbar$.
For our weak links it is of the order of $\SI{10}{GHz}$, several orders of magnitude larger than the frequencies used in our experiments. 
The RCSJ model also predicts the switching rate to be smaller than the retrapping one, inconsistent with experimental observations (see additional discussion in Supplemental Material).
We therefore conclude that our experimental  observations require a mechanism beyond the RCSJ model to explain the switching and retrapping charateristics.

\begin{figure}
\includegraphics[width=0.5\textwidth]{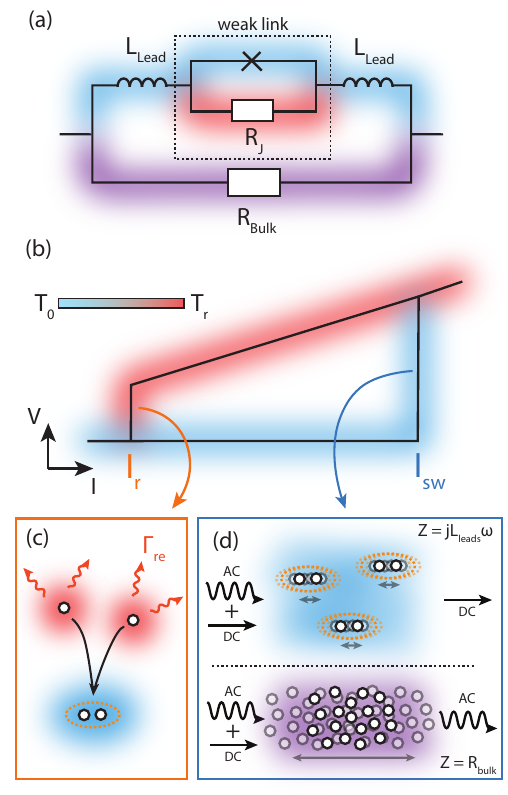}
\caption{
(a) Equivalent scheme of the MATBG junction for densities inside the lower flat band: the weak-link region modelled as a resistively shunted junction is coupled in series with the kinetic inductance of the leads. The superconducting regions (blue and red) are further shunted by normal regions (purple).
(b) Illustration of switching and retrapping mechanism and hysteresis origin in MATBG junctions. The retrapping branch of IV characteristic (red) is characterized by an increased electronic temperature $T_r$ due to Joule heating, suppressing the critical current. Retrapping into the superconducting state requires cooling the electrons (c) to base temperature characterized by a rate dependent on electronic cooling power $\Gth$.
Switching rate (blue), on the other hand, is only limited by the shunting kinetic inductance of the bulk MATBG (a,d). (c) Electronic thermal relaxation in MATBG occurs via coupling to acoustic phonons. Two electronic quasiparticles in the junction release their thermal energy to the phonon bath and become cold enough to mediate Josephson coupling.
(d) Due to their inertia, Cooper pairs in a thin superconductor (blue region) prevent the transmission of RF signals at high frequencies. Instead, the AC current at frequencies above $\omega_L =\frac{R_{bulk}}{L}$ is rerouted through non-superconducting regions of the sample (purple region) and does not affect the junction.}
\label{fig:2}
\end{figure}

\begin{figure*}[t!]
%\centering
\includegraphics[width=1\textwidth]{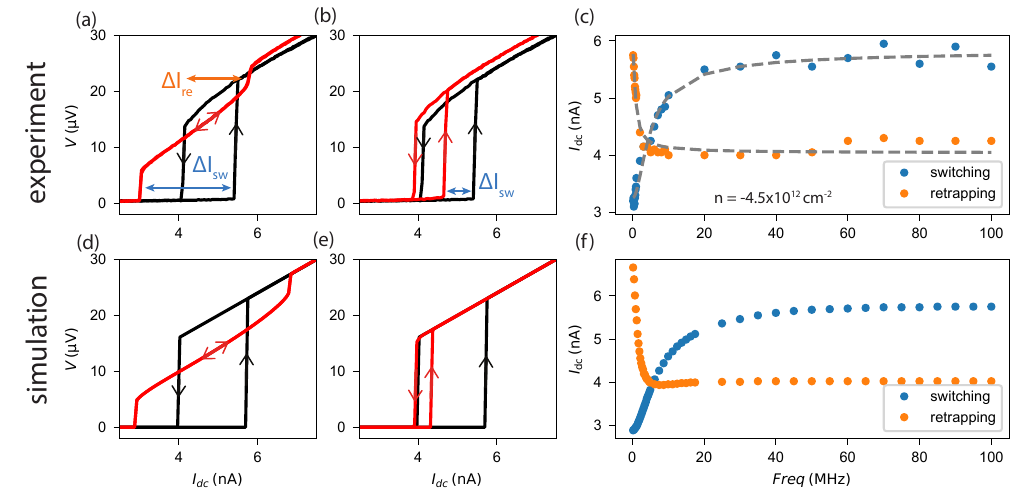}
\caption{(a) I/V traces of the junction at bias frequencies of $\SI{0.1}{MHz}$ (red) and $\SI{100}{MHz}$ (black) in the regime where the effective AC amplitude is higher than the hysteresis. (b) I/V traces of the junction at bias frequencies of $\SI{8}{MHz}$ (red) and $\SI{100}{MHz}$ (black) in the regime where the effective AC amplitude is higher than the hysteresis. The mismatch in retrapping current between the red a black curves is probably due to a charge jump (note it is of the order of a few $\mathrm{pA}$). (c) Switching and retrapping currents as a function of AC bias frequency. (d-f) Numerical simulations of our device in the same regime as the data shown in (a-c). The grey dashed line in (c) is a fit to the functional forms provided in equations \eqref{eq:RSJb} - \eqref{eq:RSJa}.
}
\label{fig:3}
\end{figure*}

%Thermal overheating explains retrapping, but not switching
Such an alternative mechanism, for both the retrapping and the hysteresis in metallic weak links is the heating of the electrons in the junction, followed by their thermalization~\cite{Courtois2008, Chiodi2008}.
In this case the retrapping branch at $I<\Isw$ is characterized by a higher temperature than the switching one due to the Joule heating in the resistive state (Figs.~\ref{fig:2} (a, b)).
This overheating reduces the weak link critical current for the retrapping branch, leading to a hysteresis.
Most importantly, retrapping back into the superconducting state requires the electronic temperature to equilibrate to base temperature, a process, depicted in Fig.~\ref{fig:2}(c), that has been directly demonstrated in superconductor-normal metal-superconductor junctions~\cite{Chiodi2008}.

%Dissipation happens through phonons.
While there are several mechanisms for energy dissipation in graphene, at low temperatures the dominant one is the coupling between electrons and acoustic phonons. 
In particular, thermalization can occur via diffusion of hot electrons into the leads, emission of blackbody photons or interaction of electrons with acoustic phonons (as the optical ones are frozen out) \cite{fong2012}. 
The first mechanism is suppressed by the presence of a superconducting gap \cite{angers2008} in the leads in our case, while the second one has been estimated to be negligible in MATBG \cite{fong2013,seifert2020magic}.
This suggests that the dominant heat loss mechanism is via coupling to phonons, in agreement with conventional SNS junctions \cite{Courtois2008, Chiodi2008}.

%By probing bulk dynamics we conclude switching is not governed by the junctions
The above mechanism on its own, however, still implies that \textit{switching} occurs with the Josephson rate $2e \Delta V/\hbar$, which is inconsistent with our observations, as detailed above.
To understand the switching dynamics in our devices we now turn to the case without a central gate voltage, i.e where the sample is homogeneously superconducting at the optimal density.
We observe a frequency-dependent IV characteristic (Fig.~\ref{fig:1}(f)), despite the absence of a weak link.
Note that there is no hysteresis, ruling out overheating as its origin.

% Switching is governed by kinetic inductance.
In addition to these observations, it has been shown that a supercurrent can flow in MATBG in narrow superconducting paths separated by normal regions \cite{Uri2020}.
The normal region thus forms a resistive shunt $\Rbulk$ coupled in parallel to the superconducting regions (purple shaded path in Fig.~\ref{fig:2} (a)).
At a non-zero frequency $\omega$, the superfluid impedance is purely inductive due to the inertia of the Cooper pairs (blue shaded mechanism in Fig.~\ref{fig:2}(d)) and given by $Z_{\mathrm{sc}} = j\omega \Lkin$, with the kinetic inductance $\Lkin \propto \frac{m^*}{\ns e^2}$, where $m^*$ is the effective mass, $e$ the electron charge, and $\ns$ is the superfluid density.
At frequencies larger than $\frac{\Rbulk}{\Lkin}$, the impedance of the superconducting branch becomes higher than the resistance of the normal bulk and the AC current flows through the non-superconducting regions (purple shaded mechanism in Fig.~\ref{fig:2}(d)).
Intriguingly, $\Lkin$ in MATBG is expected to be large \cite{Portolés2022} due to two unique properties: extremely low electron densities, and high effective mass \cite{Cao2018_1}.
This explains our observation of a rather low characteristic switching rate in Fig.~\ref{fig:1}(f).
The same mechanism applies for MATBG weak links - the kinetic inductance of bulk MATBG is then coupled in series to the junction (Fig.~\ref{fig:2}(a)).

\begin{figure*}[t!]
%\centering
\includegraphics[width=1\textwidth]{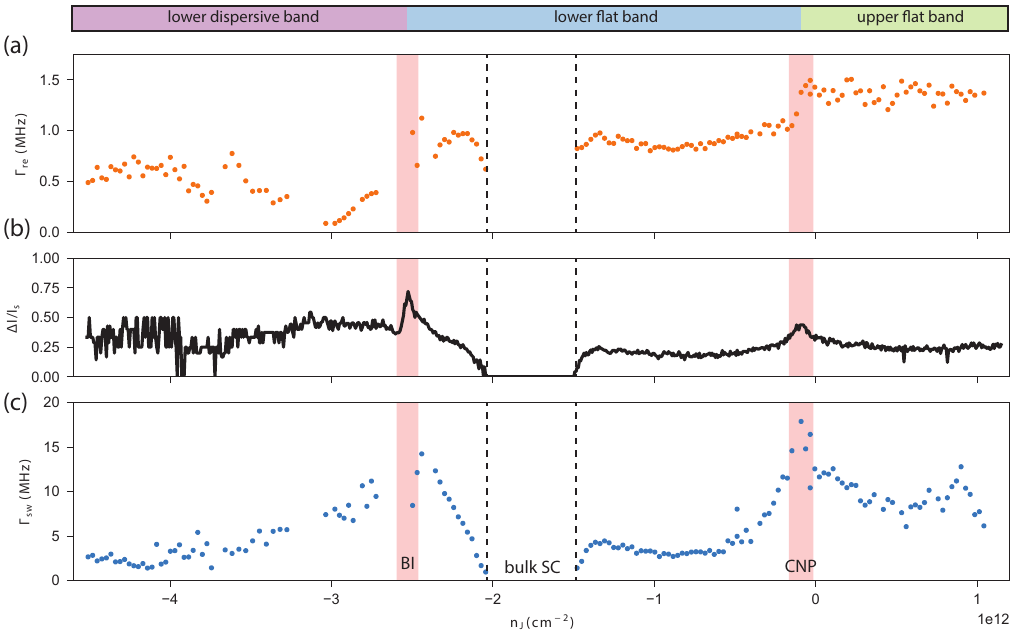}
\caption{(a) Retrapping rate as a function of junction density. (b) Relative hysteresis as a function of junction density. We observe a peak at the charge neutrality point and another one at the band insulator between lower dispersive and flat bands. (c) Switching rate as a function of junction density. The red shaded areas highlight the regions in density where a transition between bands takes place. They correspond to the Charge Neutrality Point (CNP) and Band Insulator (BI).}
\label{fig:4}
\end{figure*}

\begin{figure}[t!]
%\centering
\includegraphics[width=0.5\textwidth]{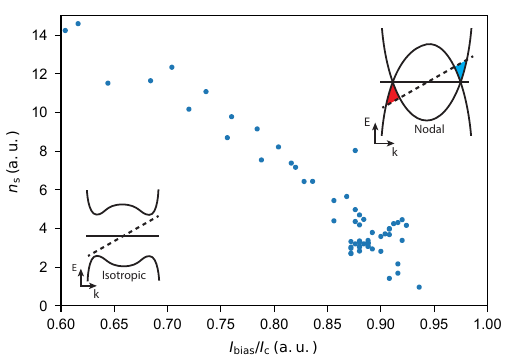}
% Change the subscripts of the quantities according to the main text
\caption{Superconducting stiffness in arbitrary units as a function of bias current to critical current ratio.
Inset, bottom left: Schematics of Bogoliubov-de Gennes quasiparticle band structure of a superconductor with isotropic gap. The solid straight line represents the Fermi energy at zero current bias. The dashed line represents the Fermi energy at a non-zero current bias. Inset, top right: Schematics of quasiparticle band structure of a superconductor with anisotropic gap. The red and blue areas represent respectively the hole and electron pockets that form at the band edges under a finite current bias. 
}
\label{fig:5}
\end{figure}

%Description of the circuit diagram
Using the ideas outlined above, we construct a model to describe the non-equilibrium dynamics of the Josephson junction.
Importantly, this model allows us to relate the observed switching and retrapping rates, $\Gammasw$ and $\Gammare$, to the microscopic and thermodynamic properties of MATBG.
The dynamics of the current-biased junction is described by:
\begin{equation}
    \Isc(t) - \Iex = \Ij (T) \sin(\varphi) + \frac{\hbar\dot{\varphi}}{2e \Rj}
\label{eq:RSJb}
\end{equation}

\begin{equation}
    \Cel \dot{T} = \frac{1}{\Rj}\left( \frac{\hbar \dot{\varphi}}{2e} \right)^2 -\Gth T
\label{eq:thermal_balance}
\end{equation}

\begin{equation}
I(t) - \Iex = \Isc - \Iex + \frac{\Lkin \dot{I}_{sc} + \frac{\hbar \dot{\varphi}}{2e}}{\Rbulk}
\label{eq:RSJa}
\end{equation}

%Description of equation 1
Equation \eqref{eq:RSJb} describes a Josephson junction with a phase difference $\varphi$, a temperature-dependent critical current $\Ij(T)$, and a fixed excess current value $\Iex$ shunted by resistance $\Rj$ (Fig.~\ref{fig:2}(a), dashed box). 
For results in the main text we assume $\Rj \ll R_{bulk}$, the general case is discussed in Supplementary Material.
We note that the form of $\Ij(T)$ has not been determined experimentally; we assume that it is a decreasing function of temperature with a single characteristic scale $\Tj$ that can be estimated to be of the order $\SI{0.1}{K}$ based on the disappearance of interference in SQUID devices \cite{Portolés2022}.
In the main text, we focus on an empirical model $\Ij = \Ij(0) \sqrt{1-T/\Tj} \cdot\theta(1-T/\Tj)$ that correctly captures the high-frequency asymptotic behavior of the retrapping current; we provide a discussion of different models and their general properties in the Supplemental Material.

%Description of equation 2
Equation \eqref{eq:thermal_balance} describes the evolution of the electronic temperature $T$ with respect to the base temperature.
The left-hand side represents the total power dissipated in the link, $\Cel$ being the electronic heat capacity.
On the right-hand side, the first term corresponds to Joule heating, while the second one is the electronic heat loss ($G_{\mathrm{th}}$) attributed, as discussed above, to electron-phonon interactions.
The processes relevant for the description of the Josephson effect occur at $T\approx \Tj$ (see Supplemental Material), such that the value of the thermal conductivity $G_{th}$ can be approximated by its value at $T=\Tj$.
The final equation describes the shunting of the junction by the resistive quasiparticles of bulk MATBG (Fig.~\ref{fig:2} (a,d)). The current $I_{sc}(t)$ is the full external current driven through the weak link. 

%Our model captures the experimental data nicely
Remarkably, we find that the model defined by Eq. \eqref{eq:RSJb}-\eqref{eq:RSJa} captures all of the behaviors observed in the experiment.
As an example, we consider a highly nonlinear regime where the RF amplitude is larger than the hysteresis $\Isw-\Ire$.
For a range of DC bias values the junction spends part of the AC period in the resistive regime and part of it being superconducting, resulting in a double step in voltage, as shown in Fig.~\ref{fig:3}(a) \footnote{Such voltage values are the average between the resistive and superconducting voltages weighted by the percentage of the time spent by the junction in each regime.}.
Fig.~\ref{fig:3}(d) shows a simulated trace in the same regime, demonstrating remarkable agreement between the model and the experiment.
As we increase the frequency of the current bias across the junction we recover the regular hysteresis (Fig.~\ref{fig:3}(a,b), black line).
The model captures the evolution of the I/V traces as the bias frequency increases, as is shown in Fig.~\ref{fig:3}(e).
Even finer details of the experimental data~\footnote{Because such details are not central to the extraction of timescales from the data, we discuss them in the Supplemental Material.}, shown in the Supplemental Material are captured by the model.
These comparisons confirm that our model accurately describes the dynamics of our junction.

%Extracting gammare and gammasw for a fixed density
To extract the retrapping and switching rates, $\Gammare$ and $\Gammasw$, for a given density from the experimental data, we fit the evolution of the retrapping and switching currents as a function of bias frequency. 
An analysis of the data, discussed in the Supplemental Material, demonstrates that both currents asymptotically approach a constant high-frequency value as $1/\omega$.
To fit the results at all frequencies, we use the following functions: 
$\Isw(\omega) = I_{\mathrm{sw},\infty}-I_{RF} \Gammasw / \sqrt{\Gammasw^2 + \omega^2}$
and $\Ire(\omega) =I_{\mathrm{re},\infty}+ I_{RF} \Gammare / \sqrt{\Gammare^2 + \omega^2}$. 
That allows to characterize the corresponding rates (see Fig.~\ref{fig:3}(c), gray lines).
The model  described in Eqs. (\ref{eq:RSJb},\ref{eq:thermal_balance},\ref{eq:RSJa}), reproduces correctly the asymptotic behavior of the switching current, while for the retrapping current the result depends on the particular form of $\Ij(T)$ (see SM).
For a fixed density in the junction, we extract the switching and retrapping currents for all frequencies and fit the results.
In the example shown in Fig.~\ref{fig:3}(c), for a density of $\SI{-4.5e12}{cm^{-2}}$, we extract $\Gammare = \SI{0.52}{MHz}$ and $\Gammasw = \SI{2.75}{MHz}$.
Therefore, the weak-link dynamics of our junction gives us access to the quasiparticle thermalization rate and kinetic inductance of MATBG (Fig.~\ref{fig:2} (c,d)).

%If we have switching rate, we have superfluid stiffness
We now provide a physical interpretation of these rates that allows us to connect them to the properties of MATBG.
We begin with the switching rate $\Gammasw$.
From Eq.~\eqref{eq:RSJa} we identify the switching rate as $\Gammasw = \Rbulk/\Lkin \propto \ns$ (see additional discussion in Supplemental Material).
Assuming that the resistance of normal regions $\Rbulk$ does not strongly depend on $T$ or bias strength,  $\Gammasw^{-1} \propto \Lkin$, which allows to probe the superfluid stiffness of MATBG.

%Important to take into account the gammare/gammasw ratio when interpreting gammare
Before discussing the thermalization rate of the weak link, we  note that 
for $\omega\gg\Gammasw$ the AC part of the current does not reach the junction at all: $I_{sc} \approx I_{DC}$. 
Thus, for $\Gammare > \Gammasw$, the kinetic inductance would set the rate for both switching and retrapping.
However, as shown in Fig.~\ref{fig:4}, we have $\Gammare$ \textit{strictly} smaller than $\Gammasw$ for all densities (note the different y-axis in Figs.~\ref{fig:4}(a,c)), confirming that we can interpret the former as a thermalization rate.

%Retrapping rate leads to Gth and Cel
The equation governing thermalization in the device in Eq.~\eqref{eq:thermal_balance} contains two implicit frequency scales: $\gamma \equiv \frac{\Gth}{\Cel}$ and $k\equiv \frac{\Ij^2 R_J}{\Cel \Tj}$.
Importantly, the hysteresis size for DC driving depends on their ratio $\gamma/k$, while the retrapping rate $\Gammare$ depends on both, allowing in principle, to determine both scales, and, as a result $\Cel$ and $\Gth$.
While the quantitative nature of $\Gammare$ depends on the particular form of $\Ij(T)$, for the square-root model introduced above and $\gamma/k<1/2$ we obtain an analytical result for the retrapping rate: $\Gammare= \frac{k \sin \frac{2 \pi \gamma}{k}}{2 \pi}$.
Furthermore, the ratio between DC retrapping current and switching is $I_r/I_s = \sqrt{\gamma}{k}$ (see Supplemental Material).
We stress that the observed $I_r$ and $I_s$ are rather close to one another, which results in $\gamma$ and $k$ being effectively of the same order of magnitude.
For larger values of $\gamma$ the model predicts an $1/\omega^2$ dependence of the retrapping current under AC bias; therefore the model should not be applicable for $\Ire/\Isw > 1/\sqrt{2}$.
However, in the absence of direct measurements of $\Ij(T)$, we will use this model to estimate $\Cel$ and $\Gth$.

%Gammare gives indeed thermal properties. The corresponding e-ph coupling is bigger that that of Aluminium.
We observe, across the whole density range, three sets of values of $\Gammare$: $\SI{0.5}{MHz}$, $\SI{1}{MHz}$ and $\SI{1.5}{MHz}$, corresponding to the chemical potential of the link tuned to the dispersive band, lower flat band and upper flat band, respectively.
The change in the hysteresis width, Fig.~\ref{fig:4}(b) is relatively smaller.
Using the analytical formula given above, we can estimate for $\Delta I/\Isw \approx 0.5$ that $\gamma \approx 0.8 - 2.3$ MHz.

This result already provides an important insight into the low-temperature behavior of electron-phonon coupling in MATBG when contrasted with those at higher temperatures.
In particular, the cooling rate has been found to be of the order of hundreds of $\mathrm{GHz}$ above $\SI{5}{K}$ with a very weak temperature dependence \cite{mehew2024ultrafast}, attributed to effective moir\'e Umklapp scattering \cite{ishizuka2021purcell} explaining the linear-in-temperature resistivity \cite{sarma2019,ishizuka2021purcell,polshyn2019large}. 
The strong difference with our result at $T\sim T_J \approx \SI{100}{mK}$ suggests a suppression of the cooling rate much stronger than linear-in-temperature. 
This result is consistent with electron-phonon scattering at 100 mK being in the Bloch-Gruneisen regime where Umklapp scattering is suppressed \cite{ishizuka2021purcell} and resistivity from electron-phonon scattering should follow a stronger power-law dependence on temperature \cite{sarma2022}.
This excludes electron-phonon scattering as the origin of linear-in-temperature resistance at low temperatures \cite{jaoui2022quantum}.

% Low electron-phonon coupling.
In the case of superconductivity, the most relevant quantity when discussing electron-phonon coupling is the dimensionless coupling constant, which we note here $\lambda$. 
The temperature relaxation rate at low temperatures is related to the strength of the coupling to acoustic phonons \cite{allen1987,viljas2010,fong2017}. 
While this coupling does not take the contribution of optical phonons into account, it is expected to be of the same order of magnitude as the full coupling constant \cite{allen1987}. 
To obtain an estimate we use a Dirac electron model \cite{sarma2019,viljas2010,fong2017}: one finds that $\gamma = \frac{\Gth}{\Cel} = \lambda \frac{16 \pi^2}{5} \frac{(k_B T_{el})^2}{\hbar^2 s k_F}$, where $s$ is the acoustic phonon velocity. 
Using $T_{el}\sim \Tj\sim \SI{0.1}{K}$ from the extinction temperature of SQUID oscillations~\cite{Portolés2022}, $s\approx 20$ km/sec (the value for single-layer graphene \cite{cong2019probing} is expected to be close to that in MATBG \cite{koshino2019}), $k_F  = \sqrt{\pi n}$ for $n\sim \SI{1e-12}{cm^{-2}}$ and $\gamma\sim 1$ MHz we obtain $\lambda \sim 10^{-3}$. 
The resulting estimate for the dimensionless coupling constant is more than an order of magnitude lower than in conventional superconductors \cite{poole1999handbook} and theoretical estimates for MATBG \cite{sarma2019}, which should be considered when discussing the mechanism behind superconductivity in MATBG.

%We can even extract more estimates.
We can further estimate $\Gth$ and $\Cel$ taking $\Ij-I_{exc}\sim \SI{5}{nA}$, $\Delta V\sim \SI{20}{\mu V}$ from Fig.~\ref{fig:3}.
The result is $G_{th}\sim \SI{250}{fW/K}$ and $\Cel\sim \SI{5e-19}{J/K}$.
From the junction area and $n\sim \SI{1e-12}{cm^{-2}}$ one expects above $10^3$ electrons, with the usual Sommerfeld expression $\Cel = \frac{\pi^2}{2}k N \frac{k_B T}{E_F} \sim 10^{-19} \frac{k_B T}{E_F}$ J/K.
In usual metals, $\frac{k_B T}{E_F}\ll1$, while in our case this implies $\frac{k_B T}{E_F} \sim 1$, that may be related to large residual entropy of interacting states of MATBG \cite{rozen2021entropic}. 
Both $\Gth$ and $\Cel$ are much higher than those expected in monolayer graphene \cite{fong2017}, consistent with strongly suppressed bandwidth and electron velocities of MATBG $G\propto v_F^{-2}, C\propto v_F$. However, the $\gamma = \frac{\Gth}{\Cel}$ we find in MATBG at $T\approx T_J$ are of the same order as those predicted for monolayer graphene \cite{fong2017}. 
Using the Dirac electron prediction $\gamma \propto D^2/v_F$, where $D$ is the deformation potential, this suggests that the deformation coupling to acoustic modes is reduced in MATBG compared to monolayer graphene. 
This explains the weak value of $\lambda$ estimated above, despite the strong increase in density of states expected in MATBG.

%gammasw tells us about ns in the flat band, but not outside of it
We now discuss the switching rate $\Gammasw$ (Fig.~\ref{fig:4}(c)), related to the superfluid stiffness in the bulk of our MATBG device.
Importantly, the AC measurements are still performed at a finite DC bias, thus, our measurements reveal the superfluid density at a finite current bias, $\ns(I_{DC}) \approx \ns(\Isw)$.
Since $\ns(\Isw)$ is a decreasing function of current, the steep increase in $\Gammasw$ at the edges of the lower flat band is explained by the decreasing critical current of the junction (Fig.~\ref{fig:1} (d)).
On the contrary, the decrease of $\Gammasw$ for densities in the top flat and dispersive bands, is unexpected - at such low critical currents $\ns(\Isw)\approx \ns(0)$ should be density-independent and large. 
We suggest that this observation can be explained by the kinetic inductance of proximity-induced superconductivity in the junction region. 
Being very weak, the proximity-induced superfluid has an extremely large kinetic inductance that is in parallel to the smaller one from the bulk TBG, effectively shunting it\footnote{The proximity effect through a junction is related to its excess current. We discuss such relation in our device in Supplementary Material.}.

%\ns vs I_bias, points to nodal
Let us now return to the densities within the lower flat band, where $\Gammasw$ is related to the superfluid density of bulk MATBG.
The dependence of $\Gammasw \propto \ns$ as a function of $I_{DC}$, shown in Fig.~\ref{fig:4}(c), gives important information about the nature of the superconducting gap in MATBG.
Current biasing a superconductor produces a Doppler shift \cite{Yip1992, Volovik1993} of the quasiparticle bands in a superconductor, see inset of Fig.~\ref{fig:5} ($\Delta E = {\bf v}_s \times{\bf \hbar \bf k}$, where ${\bf v}_s$ is the superfluid velocity and ${\bf k}$ the quasiparticle momentum).
For an isotropic superconductor, depicted in the inset of Fig.~\ref{fig:5}, this does not affect the quasiparticle occupations until a critical value of bias current is reached.
As a result, $\ns(I)$ dependence is highly nonlinear with an abrupt drop close to the critical current \cite{clemkogan}.
For a highly anisotropic or nodal superconductor, across its nodal axis in real space, the quasiparticle band structure presents cones instead of a gap in density (Fig.~\ref{fig:5}, inset).
A small shift originating from a finite bias current, leads to a finite generation of quasiparticle pairs, thus reducing the superfluid density before breaking down the superconducting condensate.
As Fig.~\ref{fig:5} shows, the relation between superfluid density and bias current is linear in the case of MATBG in the range $I_{\mathrm{dc}} \in [0.6\Ic, 0.95\Ic]$.
This result is inconsistent with the behavior expected of an isotropic superconducting gap, ruling in favor of a highly anisotropic or nodal pairing state in MATBG.

%Conclusion
In conclusion, we have presented a method for characterizing electron dynamics in twisted bilayer graphene by combining electrostatic and radiofrequency current bias of the material.
Together with a theoretical model, this allows us to relate the frequency dependence of the current-voltage characteristics of a Josephson junction to physical properties of the electrons in the material for a wide range of densities.
From the extracted electron thermalization rates we estimated the electron-phonon coupling in twisted bilayer graphene to be too weak to explain either superconductivity \cite{sarma2019,Biao2019, WuIvar-2018, Kennes2018} or strange metal behavior at low temperatures \cite{jaoui2022quantum,sarma2022}.
From the evolution of the switching current of the junction, we extracted a current bias dependence of the superfluid stiffness that points towards the superconducting gap of the material being anisotropic. 
The technique we developed in this work can be applied to a wide range of gate-tunable superconducting 2D materials, introducing a general way to access important thermodynamic quantities, such as specific heat and superfluid stiffness. 
In addition to being a valuable addition to experimental probes of 2D materials, we demonstrated a controllable driving of a correlated electronic system, opening the path to realization of out of equilibrium states of electrons.
\clearpage
\newpage

\section{Data availability}

The data that support the findings of this study will be made available online through the ETH Research Collection.

\section{Acknowledgements}
We thank Peter M\"{a}rki, Wister Huang and the staff of the ETH cleanroom facility FIRST for technical support. We thank Landry Bretheau for helpful and detailed discussions on  radiofrequency biasing of superconducting devices and members of the quantum e-leaps consortium for comments on our data.
We acknowledge financial support by the European Graphene Flagship Core3 Project, H2020 European Research Council (ERC) Synergy Grant under Grant Agreement 951541, the European Union’s Horizon 2020 research and innovation program under grant agreement number 862660/QUANTUM E LEAPS, the European Innovation Council under grant agreement number 101046231/FantastiCOF, NCCR QSIT (Swiss National Science Foundation, grant number 51NF40-185902).
K.W. and T.T. acknowledge support from the JSPS KAKENHI (Grant Numbers 21H05233 and 23H02052) and World Premier International Research Center Initiative (WPI), MEXT, Japan.
E.P. acknowledges support of a fellowship from ”la Caixa” Foundation (ID 100010434) under fellowship code LCF/BQ/EU19/11710062. 
J.H.P. acknowledges NSF Career Grant No. DMR-1941569.
This work was initiated and partly performed at the Aspen Center for Physics, which is supported by National Science Foundation grant PHY-2210452.

\section*{Author information}

\subsection*{Corresponding authors}
Correspondence and requests for materials should be addressed to E.P. or P.A.V.

\subsection*{Author contributions}
P.R. and E.P. fabricated the device.
T.T. and K.W. supplied the hBN crystals.
E.P. and M.P. performed the measurements.
E.P., P.A.V., M.P., M.T. and Y.K. analyzed the data.
P.A.V. developed the theoretical model and performed the numerical and analytical calculations.
K.E., T.I. and E.P. conceived and designed the experiment.
J.H.P., T.I. and K.E. supervised the project.
E.P. and P.A.V. wrote the manuscript with comments from all authors.

\subsection*{Competing interests}
The authors declare no competing financial interests.

\clearpage
\newpage
\onecolumngrid

\renewcommand{\thefigure}{S\arabic{figure}}
\setcounter{figure}{0} 

\renewcommand{\theequation}{S\arabic{equation}}
\setcounter{equation}{0} 

\renewcommand{\thesection}{\Alph{section}}

\setcounter{section}{0} 

\begin{center}
\Huge{Supplemental Material}
\end{center}

\begin{center}
\Large{Experimental Details}
\end{center}

\section{Fabrication Details and Measurement Setup}

The device being the same as the one presented in reference \cite{deVries2021} we refer the reader to the methods section of this reference for details about the fabrication and measurement setup.

In this work, there are however two modifications with respect to the aforementioned setup. The first one is that the AC bias is not sent to the central gate but to one of the leads, using a bias T to be able to send both AC and DC signals to the same contact. The other difference is that, because for the AC measurements we present in this study we need a higher degree of precision than for the ones presented in reference \cite{deVries2021}, we must ensure that the AC amplitude reaching the device is neither frequency-dependent nor sample-resistance-dependent. The details of such procedure are given in the following section.

\section{Radiofrequency Biasing}

When applying a radiofrequency bias to our junction two different aspects must be taken into account. The first one is the evolution of the amplitude as a function of its frequency. Indeed, having a frequency-dependent amplitude reaching our junction would make it impossible to disentangle such effects from the physical mechanisms taking place at the junction level. We performed simulations of our circuit using the software \textit{LTspice} and obtain an evolution in amplitude of our radiofrequency bias reaching the junction of less than $\SI{5}{\%}$, for frequencies ranging from $\SI{100}{kHz}$ to $\SI{100}{MHz}$. We thus conclude that, in comparison to the experimental results, these effects can be neglected.

Because the junction changes its state during the acquisition of an I/V trace, we must also ensure that the change in resistance triggered by the RF biasing does not significantly affect the AC current flowing through the device. In order to achieve that, we place a $\SI{100}{k\Omega}$ resistor in series between the RF feed line and the device. Like for the previous effect, this leads to a variation in AC amplitude reaching the device of the order of a few percents across the whole resistivity range of the device. We thus can also neglect this effect taking into account the precision of the claims made in our analysis.

\section{Frequency dependence of the retrapping current}
\begin{figure*}[h]
\includegraphics[width=0.7\textwidth]{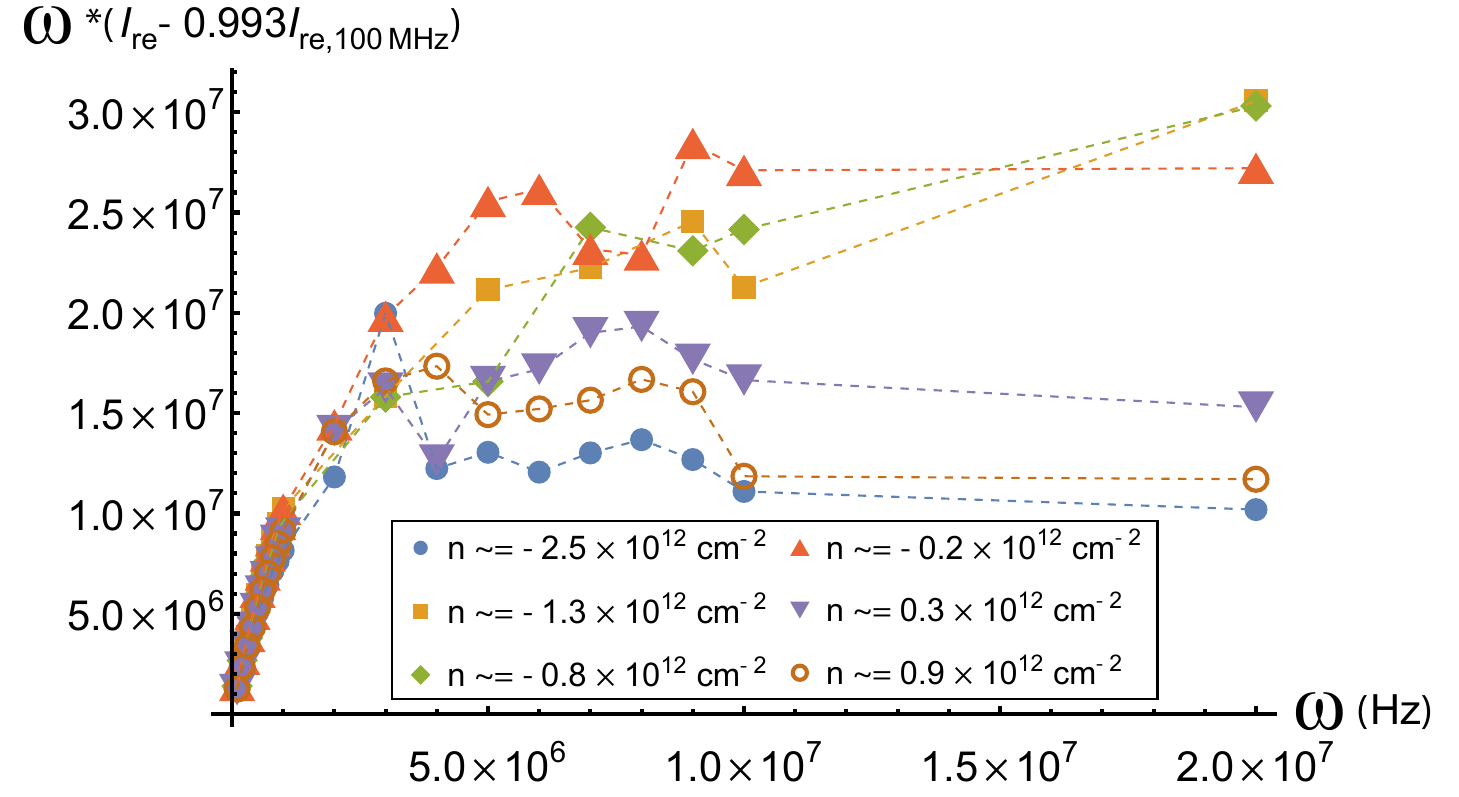}
\caption{Asymptotic behavior of the retrapping current on frequency (lines are guide to the eye). $[\Ire-0.993\Ire(100\; MHz)]\omega$ is approximately constant above 5 MHz for a range of electron concentrations suggesting a $1/\omega$ asymptotic behavior.
}
\label{fig:SM_scaling}
\end{figure*}

We show that the retrapping current shows an asymptotic $1/\omega$ behavior in a wide range of densities. In Fig.~\ref{fig:SM_scaling} we show that $[\Ire-\Ire(100\; MHz)]\omega$ is approximately constant above 5 MHz for a range of electron densities. This motivates the empirical fitting function used to extract $\Gammare$ in the main text. For theoretical explanation of this behavior see below.

\section{Hysteretic double step}

\begin{figure*}[h]
\includegraphics[width=0.5\textwidth]{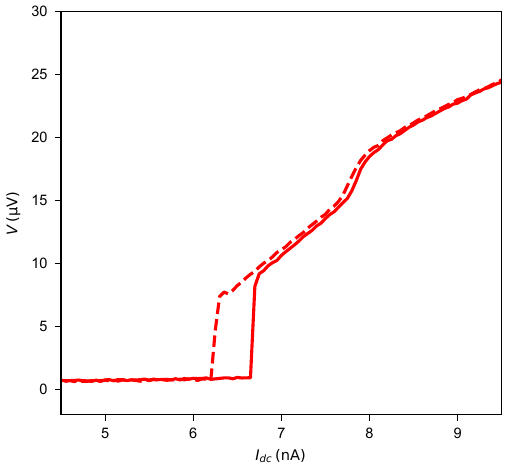}
\caption{Current-voltage characteristic of the junction for a density of $n=\SI{-4.56}{cm^{-2}}$ and an RF frequency of $\SI{800}{kHz}$. Solid line represents data from increasing current bias while dashed lines represents data from decreasing current bias.
}
\label{fig:S1}
\end{figure*}

For a narrow range of electron densities and RF frequencies, we observe a double step behavior where the low current step shows some hysteresis. We show an example in figure \ref{fig:S1}. This feature is captured by our model, as discussed in latter sections (see e.g. Fig.~\ref{fig:ivtyp} (b)).

\clearpage
\newpage

\begin{center}
\Large{Theoretical Analysis}
\end{center}

In this part of the Supplemental Material we present the theoretical analysis leading to the model, presented in the main text, Eq. (1-3) and give the details of the derivation of the results used there.
In Sec. \ref{sup:sec:models} we introduce the RCSJ and heating-induction models for the Josephson junction dynamics. In Sec. \ref{sec:rcsj} we analyze the switching and retrapping in RCSJ model and show that it is both quantitatively and qualitatively inconsistent with the experimental results reported in the main text. In Sec. \ref{sec:heat-mod} the model of junction dynamics based on overheating and kinetic inductance is studied. In particular, in Sec. \ref{sec:numerics} we present the details of numerical solutions; in Sec. \ref{sec:heat:an_dc} we discuss the DC properties of the model analytically, in Sec. \ref{sec:heat:an_ac} we study the AC properties analytically, while in 
\ref{sec:finiteR} we discuss the effects of a finite $R_J/R_{bulk}$ ratio.

\section{Considered models}
\label{sup:sec:models}

In this section we introduce the models considered to describe the IV characteristics of the MATBG junctions. In particular, we consider the IV characteristics of DC and AC-current driven Josephson junctions with two models. The first model we consider is the RCSJ model \cite{Tinkham2004,shmidt1997physics}:

\begin{equation}
\frac{\hbar C}{2e} \ddot\varphi(t) 
+
\frac{\hbar}{2eR_J} \dot \varphi 
+
I_c \sin(\varphi)
=
I(t).
\end{equation}
For theoretical analysis it can be recast in dimensionless form:
\begin{equation}
\beta_c \ddot\varphi(t) 
+
\dot \varphi 
+
\sin(\varphi)
=
i(t),
\label{eq:rcsj}
\end{equation}
where $\beta_c = \frac{2e I_c R_J^2 C}{\hbar}$ is the Stewart-McCumber parameter and time is normalized as $t\to\omega_{RSJ} t$, where $\omega_{RSJ} = \frac{2e I_c R_J}{\hbar}$. Below (see Sec. \ref{sec:rcsj}) we demonstrate that RCSJ models fails both qualitatively and quantitavely in describing the experimental result presented in the main text.

Alternatively, we use a model neglecting capacitance, but including the effects of heating and kinetic inductance $L_{kin}$ presented in Eq. (1-3) of the main text. Here we reproduce them for convenience:
\begin{equation}
        I_{sc}(t)-I_{ex} = I_J(T) \sin(\varphi) + \frac{\hbar\dot {\varphi}}{2e R_J},
         \label{eq:rsj}
\end{equation}
\begin{equation}
        \Cel \dot T =  \frac{1}{R_J}\left(\frac{\hbar\dot {\varphi}}{2e }\right)^2
        -G_{th} T,
         \label{eq:heat}
\end{equation}
\begin{equation}
        I(t)-I_{ex} = I_{sc}(t)-I_{ex} + \frac{L \dot I_{sc}+\frac{\hbar\dot {\varphi}}{2e}}{R_{bulk}}.
        \label{eq:ind}
\end{equation}
In these equations, the dynamical variables are $I_{sc}$ (current flowing through the superconducting path), $\varphi$ (phase difference across the junction) and $T$ (electronic temperature in the junction); $I_{ex}$ is the excess current (see main text), $R_J$ is the junction resistance, $\Cel$ is the electronic specific heat of the junction region, $\Gth$ is the thermal conductance for electron heat loss, $L_{kin}$ - the kinetic inductance of the superconducting path and $R_{bulk}$ is the resistance of he nonsuperconducting path across the sample.

Temperature is measured from the base temperature of the cryostat. We note that the detailed temperature dependence of the Josephson current $I_J(T)$ is not experimentally known, although a characteristic temperature of 90 mK is suggested from the temperature dependence of SQUID critical current oscillations \cite{Portolés2022}. Therefore, we will assume in what follows that $I_J(T)$ is determined by a single temperature scale, i.e. $I_J(T) = I_J(0) i_J(\tau)$, where 
\begin{equation}
\tau=T/T_J,
\end{equation}
and $i_J(\tau)$ decreases strongly for $\tau>1$.

Below we discuss some general results before focusing on two empirical models
\begin{equation}
(1):  i_J(\tau) = (1-\tau)\theta(1-\tau)
   \label{eq:ij_tdep1}
\end{equation}
\begin{equation}
(2):   i_J(\tau) = \sqrt{1-\tau}\theta(1-\tau)
    \label{eq:ij_tdep2}
\end{equation}
In addition, we considered $I_J(\tau) = \theta(1-\tau)$, but this form results generically in absence of a voltage jump at the switching current.

For numerical calculations and theoretical analysis it is convenient to use dimensionless units:

\begin{equation}
       i(t) = i_{SC}(t) + \frac{\dot i_{SC}}{\omega_L} +\dot {\varphi}\frac{R_J}{R},
        \label{eq:indnum}
\end{equation}
\begin{equation}
        i_0(t) = i_J(\tau) \sin(\varphi) + \dot {\varphi},
         \label{eq:rsjnum}
\end{equation}
\begin{equation}
    \dot \tau = - \gamma \tau + k \dot \varphi^2.
         \label{eq:heatnum}
\end{equation}

\begin{equation}
\begin{gathered}
       i(t) = i_{SC}(t) + \frac{\dot i_{SC}}{\omega_L} +\dot {\varphi}\frac{R_J}{R},
\\
        i_0(t) = i_J(\tau) \sin(\varphi) + \dot {\varphi},
\\
    \dot \tau = - \gamma \tau + k \dot \varphi^2,
    \end{gathered}
         \label{eq:num}
\end{equation}
where $\omega_L =\frac{R_{bulk}}{L}$, $\gamma=\frac{G_{th}}{C}$, $k = \frac{I_J^2(0)R_J}{C T_J}$; all currents are normalized to $I_J(0)$, all frequencies  - to 
\begin{equation}
    \omega_{RSJ}=\frac{2e I_J(0) R_N}{\hbar},
\end{equation}
all temperatures to $T_J$, i.e. $\tau=T/T_J$. Finally, $t\to \omega_{RSJ}t$.

\section{RCSJ Model}
\label{sec:rcsj}

In this Section we analyze the IV characteristics of the RCSJ model, Eq. \eqref{eq:rcsj}, and show that it is inconsistent with the experimental results. Eq.~\eqref{eq:rcsj} contains only one free parameter - $\beta_c$. It can therefore be determined from the size of the hysteresis in the IV characteristic $(I_{sw}-I_r)/I_{sw}$ and for the TBG devices under consideration we find moderate values, e.g. $\beta_c=4$ \cite{deVries2021}. In the main text, we argue that for such values of $\beta_c$, the characteristic frequencies of the system remain of the order $\omega_{RSJ}$ which is of the order GHz, grossly inconsistent with the observed values of $\Gamma_{re,sw}$ (which are of the order MHz, see Fig.~4 of the main text). Here we show that RCSJ model results in several qualitative features also inconsistent with the experimental observations.

\begin{figure}[h!]
    \centering
    \includegraphics[width=0.5\textwidth]{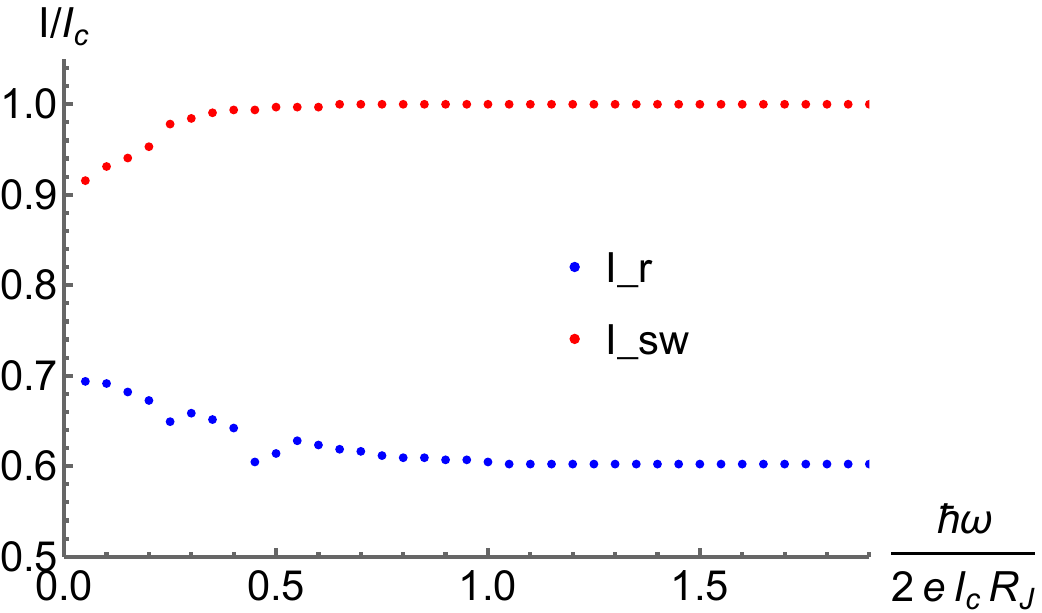}
    \caption{Dependence of switching and retrapping current on the driving frequency in the RCSJ model, Eq. \eqref{eq:rcsj}, for $i_{RF}=0.1,\beta_c=4$.
    }
    \label{fig:rcsj}
\end{figure}

In Fig.~\ref{fig:rcsj} we present $I_{sw}$ and $I_r$ as a function of RF frequency for $i_{RF}=0.1,\beta_c=4$. The results have been obtained by numerically solving Eq.\eqref{eq:rcsj} with initial conditions $\dot{\varphi}=0,\varphi(0) = \{\arcsin(i), (i<1);\;\pi/2, (i>1)\}$, where $i=I/I_J(0)$,
for the switching branch and  $\dot{\varphi}=200,\varphi(0) = 0$ for the retrapping one (large value of $\dot{\varphi}$ forces the solution into the finite-voltage state if it is stable). 

\begin{figure}[h!]
    \centering
\begin{minipage}{.45\textwidth}    \includegraphics[width=\textwidth]{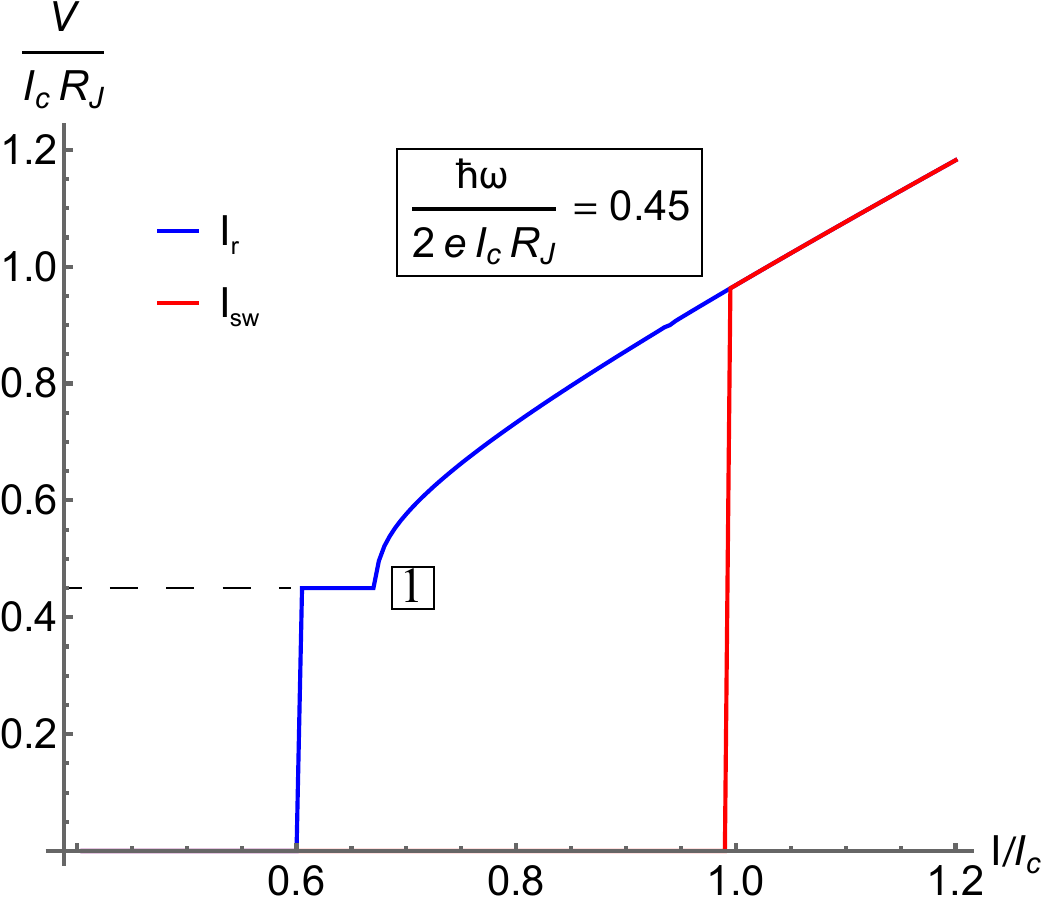}
    \end{minipage}
\begin{minipage}{.45\textwidth}
    \includegraphics[width=\textwidth]{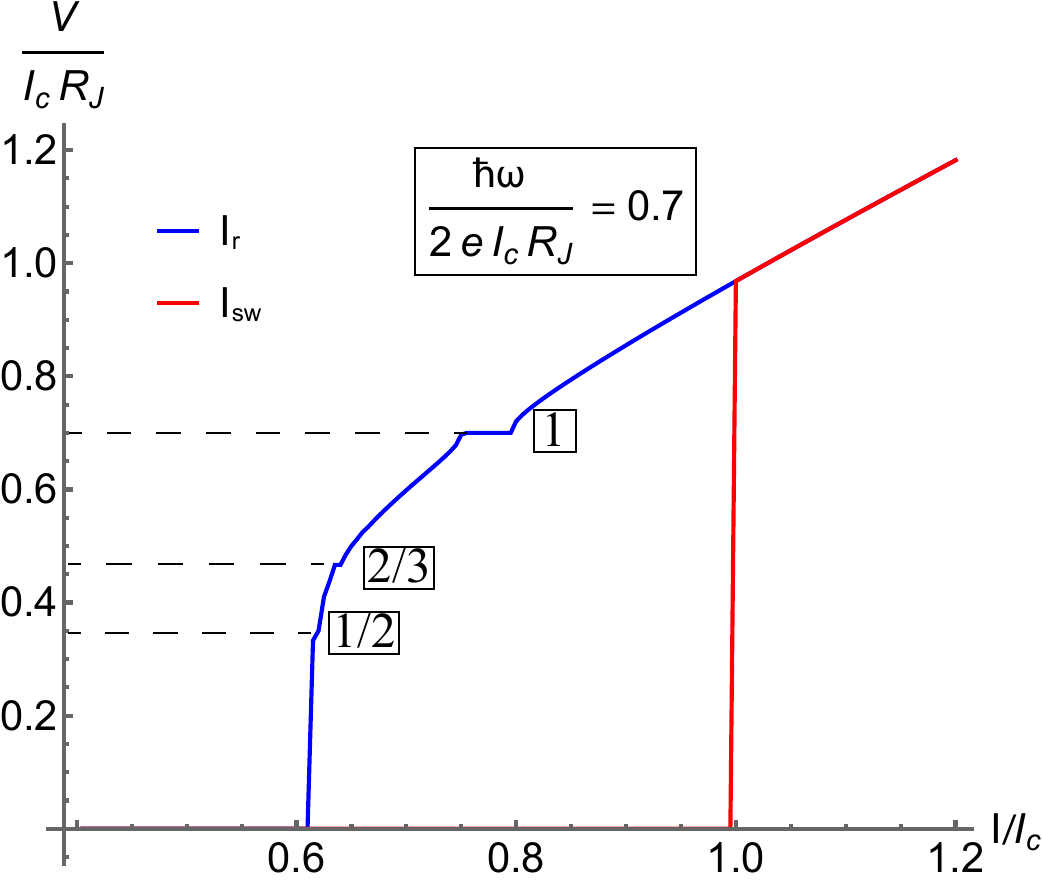}
    \end{minipage}
    \caption{Current-voltage characteristics of the RCSJ model (Fig.~\ref{fig:rcsj}) for two frequency values and $i_{RF}=0.1,\beta_c=4$. Shapiro steps are marked by black dashed lines with the corresponding fraction shown in the adjacent rectangle. Note that for higher frequency (b) fractional Shapiro steps are seen, which are expected for underdamped junctions \cite{belykh1977,bohr1984}.
    }
    \label{fig:rcsj2}
\end{figure}

One notes that both $I_{sw}$ and $I_r$ saturate at high frequency above about $0.3 \omega_{RSJ}$ for $I_{sw}$ and 
$\approx 0.7 \omega_{RSJ}$ for $I_r$. Importantly, $I_{sw}$ is recovered before $I_r$ on increasing frequency, which is inconsistent with experimental results, where the opposite is true (see Fig.~3 c and 4 of the main text).  In addition, there are additional dip-like features in $I_{sw}(\omega)$. The latter can be attributed to the development of Shapiro steps \cite{Tinkham2004,shmidt1997physics}. In Fig.~\ref{fig:rcsj2}, left panel, a Shapiro step appears right at the retrapping current, explaining the anomalous behavior in Fig.~\ref{fig:rcsj} at this frequency. Note that Shapiro steps also occur away from this frequency (Fig.~\ref{fig:rcsj2}, right panel). Therefore, both the lower characteristic frequency observed for $I_{sw}$ than that for $I_r$ and the presence of Shapiro steps are qualitatively inconsistent with experimental observations, ruling bare RCSJ model out. 

\section{Heating \& Inductance Model}
\label{sec:heat-mod}

For most of the discussion below we assume $R\gg R_J$, such that the term with $\dot{\varphi}$ in Eq. \eqref{eq:ind} drops out. For discussion of its effects see Sec. \ref{sec:finiteR}

\begin{equation}
        I(t)-I_{exc} = I_{SC}-I_{exc} + \frac{L \dot I_{SC}}{R},
        \label{eq:ind0}
\end{equation}

Additionally, without loss of generality we assume that the excess current is absorbed into $I_{SC}$ for brevity of notation.

\subsection{Numerical solutions}
\label{sec:numerics}

Here we illustrate in more detail the IV characteristics resulting from Eqs. \eqref{eq:num}. For comparison with main text we use $\omega_{RSJ} = 250$ MHz, $I_J(0)= 5.75$ nA; we note that these are simply unit choices.

\begin{figure}[h!]
    \centering
\includegraphics[width=\textwidth]{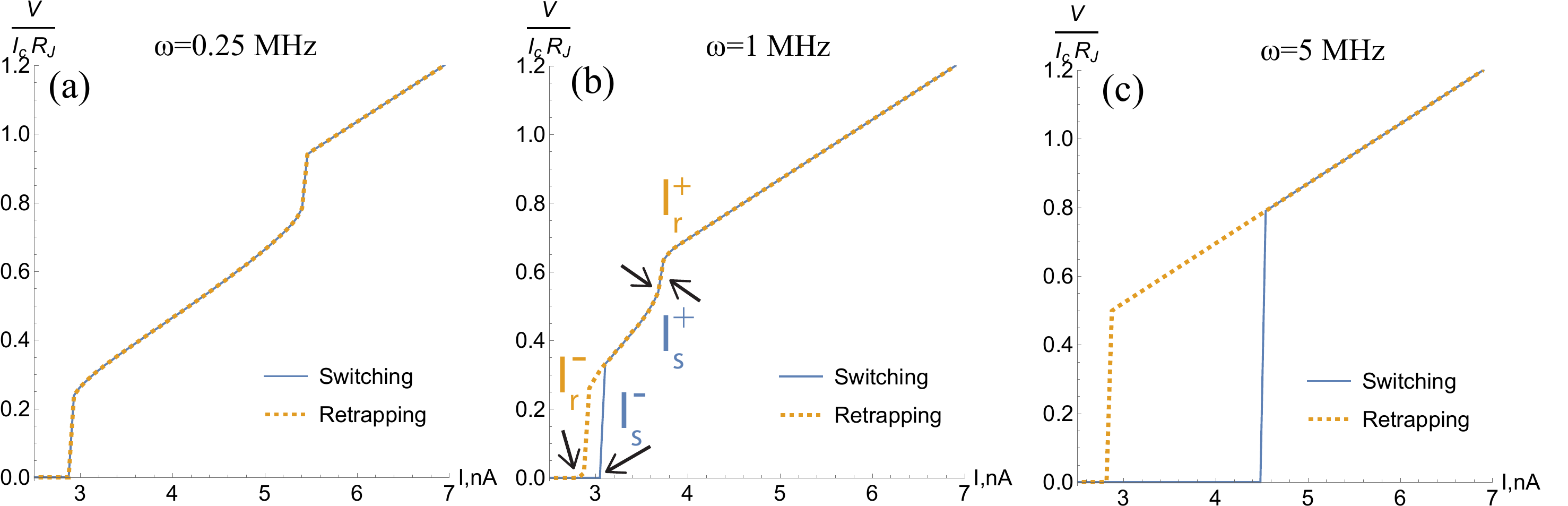}
    \caption{IV characteristics for model (2) [used in the main text],  Eqs. \eqref{eq:num} and \eqref{eq:ij_tdep2} for $k=0.01, \gamma= 0.0025, \omega_L = 2.5$ MHz, $I_{RF}/I_J(0) = 0.5$, $I_J(0) = 5.75$ nA. Four characteristic current values are introduced in middle panel, see text.}
    \label{fig:ivtyp}
\end{figure}

We begin by presenting in Fig.~\ref{fig:ivtyp} three typical IV characteristics obtained for $k=0.01, \gamma= 0.0025, I_{RF}/I_J(0) = 0.5, \omega_L = 2.5$ MHz (in Fig.~3 of the main text $\gamma= 0.0049$ is used). At small frequency (Fig.~\ref{fig:ivtyp} (a)) the IV curves are identical on switching and retrapping and show the double step behavior. Above the second feature the IV curve is purely linear. On increasing frequency (Fig.~\ref{fig:ivtyp} (b)), the lower part of the double step splits into two for switching/retrapping. This allows us to define four characteristic current values: where the IV curve deviates at high-current linear behavior ($I_r^+,I_s^+$) and where the voltage becomes zero ($I_r^-,I_s^-$). At higher frequencies (Fig.~\ref{fig:ivtyp} (c)), the splitting at the lower end of double step grows and eventually there remains only a single characteristic value for switching/retrapping.

\begin{figure}[h!]
\centering
\begin{minipage}{.3\textwidth}
  \centering
  \includegraphics[width=\linewidth]{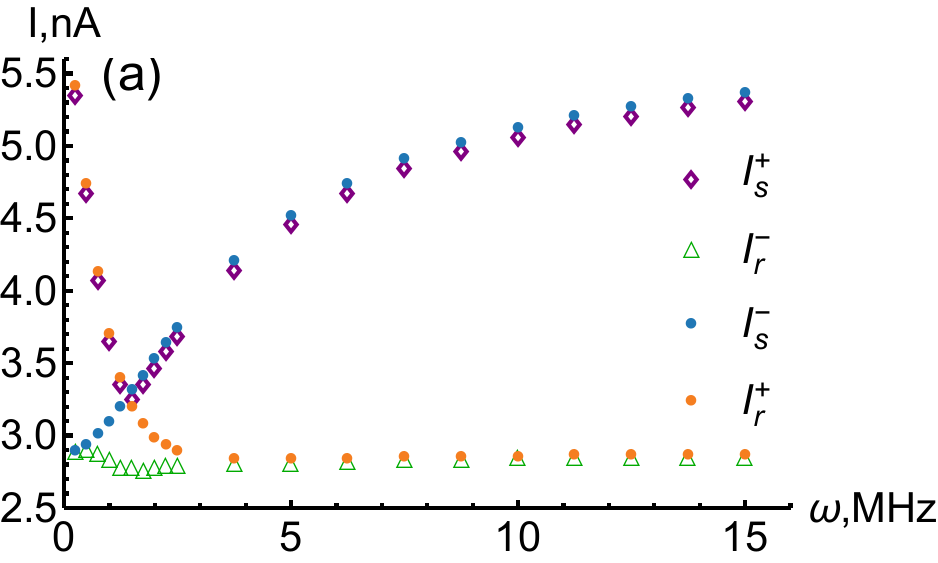}
\end{minipage}%
\begin{minipage}{.3\textwidth}
  \centering
  \includegraphics[width=\linewidth]{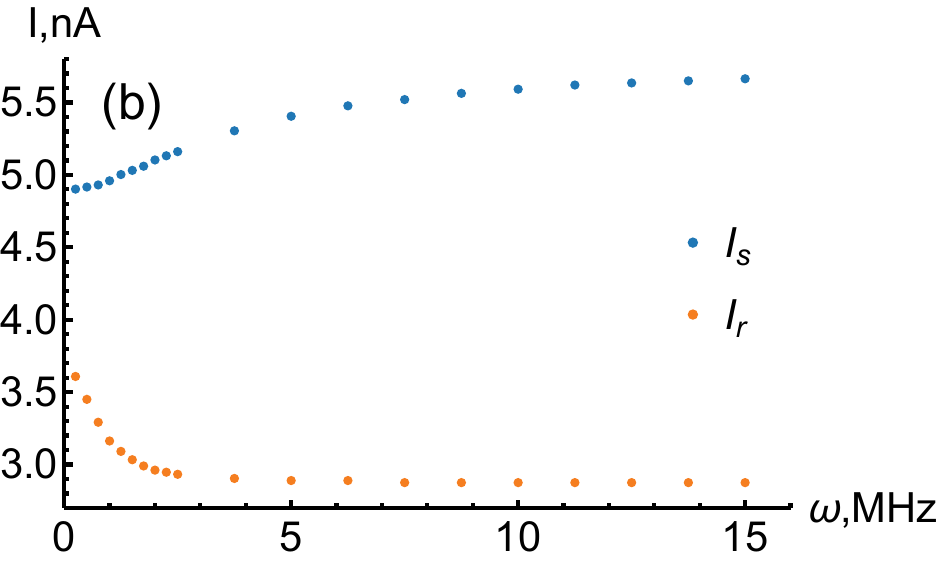}
\end{minipage}
\begin{minipage}{.3\textwidth}
  \centering
  \includegraphics[width=\linewidth]{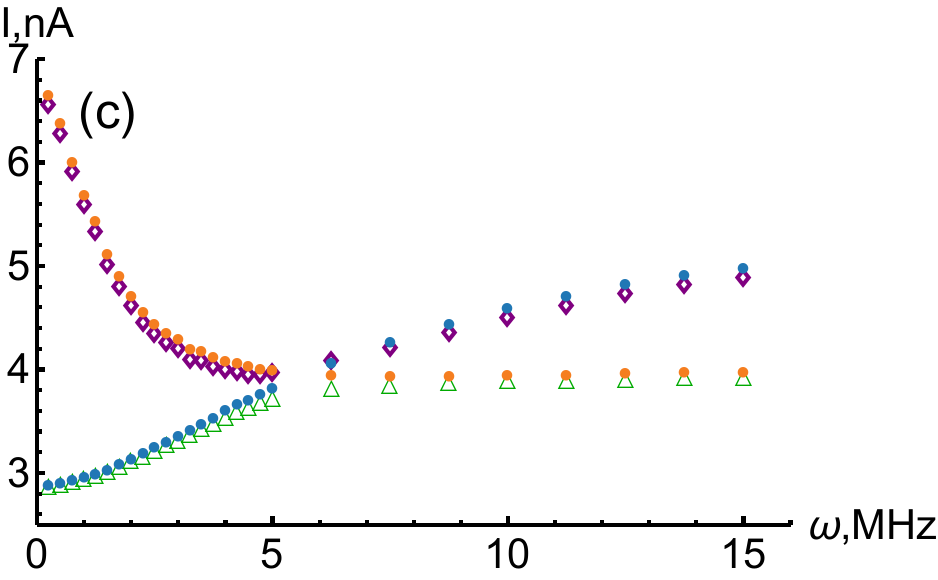}
\end{minipage}
    \caption{Frequency dependence of characteristic current values defined in Fig.~\ref{fig:ivtyp} (b) for model (2) [used in the main text],  Eqs. \eqref{eq:num} and \eqref{eq:ij_tdep2}. In  
    (a), same parameters, as in Fig.~\ref{fig:ivtyp} are used. In (b), a reduced $I_{RF}/I_J(0)=0.15$ is used,  such that there is no double step at all frequencies. In (c),  the parameters used in Fig.~3 of the main text are used.
    }
    \label{fig:freq}
\end{figure}

\begin{figure}[h!]
\centering
\begin{minipage}{.5\textwidth}
  \centering
  \includegraphics[width=\linewidth]{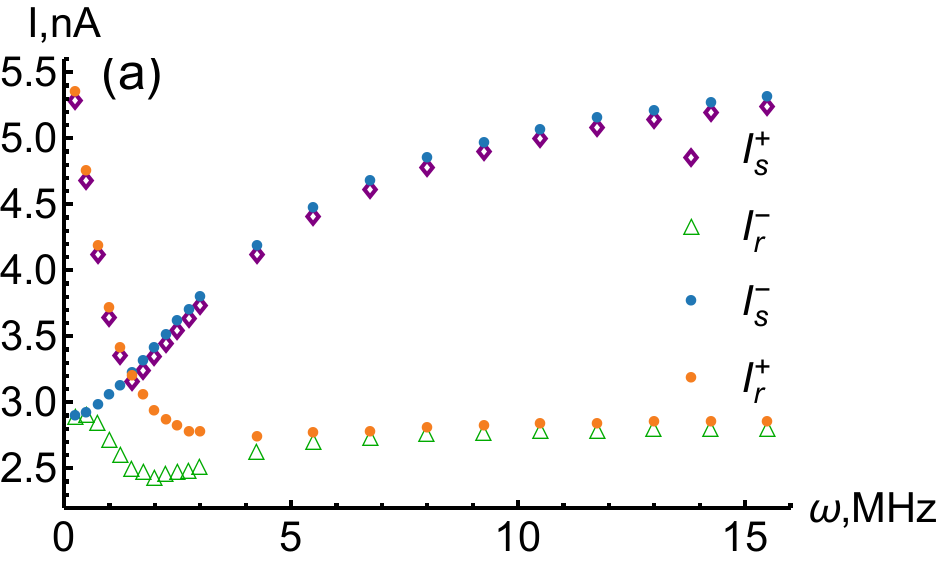}
\end{minipage}%
\begin{minipage}{.5\textwidth}
  \centering
  \includegraphics[width=\linewidth]{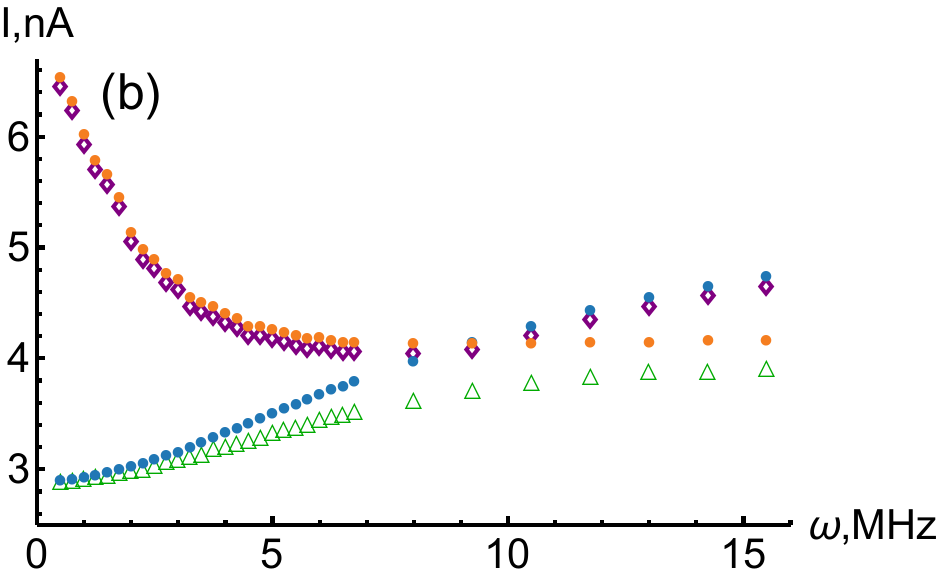}
\end{minipage}
    \caption{Same as Fig.~\ref{fig:freq} but computed using the model (1), Eqs. \eqref{eq:num} and \eqref{eq:ij_tdep2}, with(a) $\gamma= 0.0069$ and (b) $\gamma =0.0029$. While the results of the two models looks superficially similar, the asymptotic behavior at large frequencies is different (see Sec. \ref{sec:heat:an_ac}).
    }
    \label{fig:freq2}
\end{figure}

Tracking these values as a function of frequency, we obtain Fig.~\ref{fig:freq}. Importantly, the characteristic frequency values do not depend strongly on the AC current amplitude - in Fig.~\ref{fig:freq} (b) the switching and retrapping current values saturate at roughly the same frequencies as in (a) despite $I_{RF}$ being smaller such that no double step exists. 
In panel (c) we also show the results for the same parameters as in main text. The splitting of the lower edge of the double step is smaller in that case.
The main qualitative features presented above for model (2) also occur for a different choice of $I_J(T)$. In Fig.~\ref{fig:freq2} we present the characteristic current values from IV curves for model (1) with (a) $\gamma= 0.0069$ and (b) $\gamma =0.0029$ to reproduce the same hysteresis size without AC drive as with model (2).
However, as shown below, only model (2), Eq. \eqref{eq:ij_tdep2}, allows to reproduce $\propto 1/\omega$ behavior of the retrapping current observed experimentally.

\subsection{Analytics: DC case and hysteresis}

\label{sec:heat:an_dc}

We first consider the IV characteristic with a purely DC current drive. The equation \eqref{eq:ind} is reduced then to $I_{SC}-I_{exc} = I-I_{exc}$, or $i_0 = i$. 
Furthermore, the equations can be drastically simplified assuming the timescales for temperature evolution to be much longer than the RSJ scale, i.e. $\gamma, k\ll1$. Then, one can solve the RSJ equation assuming a $\tau$
\cite{shmidt1997physics}:
\begin{equation}
    \frac{\hbar\dot {\varphi}}{2e } = V(t) = R_J \frac{I_0^2-I_J^2(\tau)}{I_0+I_J(\tau) \cos( \sqrt{I_0^2-I_J^2(\tau)} \omega_{RSJ} t)}.
    \label{eq:phidotadiab}
\end{equation}
Let us assume that $\sqrt{I_0^2-I_J^2(\tau)} \omega_{RSJ}\gg \frac{G_{th}}{C} = \frac{\gamma}{k}$. In this case, the oscillatory component of $\dot \varphi(t)$ leads only to a weak variation of temperature suppressed by the factor $\frac{\gamma}{\sqrt{I_0^2-I_J^2(\tau)} \omega_{RSJ}}$ with respect to the average one. This approximation will turn out to be correct for all of the results, as is shown below. In this approximation, we can then substitute $\dot \varphi^2$ with its time average resulting in:
\begin{equation}
    \dot \tau = - \gamma \tau + k  i\sqrt{i^2-i_J^2(\tau)} \theta(i^2-i_J^2(\tau)),
    \label{eq:dc_an}
\end{equation}
where the $\theta(i^2-i_J^2(\tau))$ function reflects the absence of Joule heating in the superconducting state $i<i_J$.

\subsubsection{General properties}

Several general properties can be deduced from \eqref{eq:dc_an} with few general assumptions about the form of $i_J^2(\tau)$. In particular, (1): $i_J(0)=1$ (by normalization) and (2): $i_J(\tau>1)\ll1$ (assumes that there is a characteristic temperature scale $T_J$ for the Josephson current). In the DC dirve case, the steady-state solutions of Eq. \eqref{eq:dc_an}, i.e. $\dot\tau=0$, are experimentally observable.

For $i<1$ there is always a solution $\tau=0$, $\langle \dot \varphi\rangle = 0$, reflecting the superconducting branch of the IV characteristic.
For sufficiently large $i\gg\sqrt{\frac{\gamma}{k}}$ there also exists a stable solution $\tau = \frac{k}{\gamma} i^2 \gg 1$, $i_J(\tau\ll1) \ll i$, where we used assumption (2). This solution is characterized by a finite voltage corresponding to Ohm's law $V \approx I R_J$. Note that for $k\gg \gamma$ this solution can exist for $i<1$ and therefore coexist with the superconducting solution. This implies existence of a hysteretic IV characteristic with a discontinuous transition between the two at $i=1$ (switching).

Finally, one can show that the resistive branch typically does not extend to $i=0$, implying the existence of an abrupt transition into the superconducting state at $i_r<1$. In particular, the r.h.s. of Eq. \eqref{eq:dc_an} is simply $-\gamma \tau$ for $i<i_J(\tau)$. For small $i\ll1$, $\tau^*$, where $i=i_J(\tau^*)$ is expected to be of the order $1$. Above $\tau^*$, the r.h.s. of Eq. \eqref{eq:dc_an} can be estimated to be below $-\gamma\tau^*+ki^2$. Assuming $\tau^*\approx 1$ for $i\to 0 $, we find that for $i\ll \gamma/k$ the r.h.s. of Eq. \eqref{eq:dc_an} is negative for all $\tau$, implying that the only steady-state solution is $\tau=0$ in this regime.

\subsubsection{Switching}

We now solve the equation \eqref{eq:dc_an} for a particular form of $I_J(\tau)$ given by \eqref{eq:ij_tdep1}, \eqref{eq:ij_tdep2}. For $i=1$ in \eqref{eq:dc_an}, we obtain a steady-state solution with $\tau=0$ and $\tau=\Delta\tau$, where
\begin{equation}
    \Delta \tau = 
    \left.
    \begin{cases}
    \frac{2 g^2}{g^2+1}& g<1\\
    g& g\geq 1
    \end{cases}\right|_{\mathrm{model}\,\,\,(1)},
    \left.
    \begin{cases}
    g^2& g<1\\
    g& g\geq 1
    \end{cases}
    \right|_{\mathrm{model}\,\,\,(2)},
    \label{eq:deltau}
\end{equation}
where
\begin{equation}
g= \frac{I_c^2 R_J}{T_J G_{th}} \equiv \frac{k}{\gamma}.
\label{eq:gdef}
\end{equation}
As a result, there is a finite voltage jump at $I_0=I_c$
\begin{equation}
    \Delta V = 
       \left.
    \begin{cases}
   I_cR_J
    \frac{2 g}{g^2+1}& g<1\\
    I_c R_J& g\geq 1
    \end{cases}
    \right|_{\mathrm{model}\,\,\,(1)},
    \left.
    \begin{cases}
   I_0R_J
    g& g<1\\
    I_0 R_J& g\geq 1
    \end{cases}
      \right|_{\mathrm{model}\,\,\,(2)}
    .
\end{equation}
The case $g\geq 1$ corresponds to the Josephson current being completely suppressed by heating. The IV characteristic is expected then to be simply linear already above $I_c$: $V= I_0 R_J$. In both cases, at large enough current values one expects that the Joule heating completely suppresses $I_J$, resulting in $\frac{\dot \varphi}{2e} = I_0 R_J$ and $T = \frac{I_0^2 R_J}{G_{th}}>T_J$.

\subsubsection{Retrapping}
The solution \eqref{eq:deltau} with a finite temperature at $I_0 = I_c$ does not immediately cease to exist as $I_0$ is decreased. This indicates a bistability of the system, such that decreasing current below $I_c$ does not immediately lead to retrapping into the superconducting state. Instead, retrapping occurs when a finite voltage/finite temperature solution ceases to exist.

In particular, let us consider the steady-state solutions of equation \eqref{eq:num} with $i_J(\tau) = \sqrt{1-\tau}$ (model (2),  Eq. \eqref{eq:ij_tdep2}) for $\tau<1$
\begin{equation}
    \tau = g i \sqrt{i^2-(1-\tau)}. %, \tau<1.
\end{equation}
The solutions are:
\begin{equation}
    \tau = \frac{(g i)^2\pm gi \sqrt{g^2i^2+4i^2-4} }{2}.
    \label{eq:tempretr}
\end{equation}
The solution with "-" sign can be shown to be unstable, i.e. any small deviation from it will make the system evolve away from this value - $\tau$ will grow for $\tau=\tau_-+0$ and decrease for $\tau=\tau_--0$. At $i=1$, as noted above, $\tau_-=0$ and becomes negative for larger $i$, indicating the instability of the zero-temperature solution.

Eq. \eqref{eq:tempretr} also allows to obtain retrapping current value, where real solutions for $\tau$ cease to exist and finite $\tau$ becomes unstable. This occurs when the expression under the square root in Eq. \eqref{eq:tempretr} turns to zero; as this expression is a monotonic function of $i$, no new solutions may appear for lower $i$. The resulting value is $i_r = \frac{2}{\sqrt{g^2+4}}$, however, one notes that the resulting solution does not necessarily satisfy $\tau<1$: $\tau_r = \frac{g i_r^2}{2} = \frac{2 g^2 }{4+g^2}$. For $g>2$, $\tau_r>1$ and the above analysis is inapplicable. In that case, the maximum of the l.h.s. of Eq. \eqref{eq:dc_an} is always at $\tau=1$ for a fixed $i$. Therefore, retrapping occurs when the r.h.s. of Eq. \eqref{eq:tempretr} at $\tau=1$ becomes zero, i.e. $-\gamma +k i_r^2 = 0$. It is clear that for lower $i$ the r.h.s of Eq. \eqref{eq:tempretr} will be negative and thus only $\tau=0$ will be a steady state solution. Taking both cases into account and performing similar calculations for case Eq. \eqref{eq:ij_tdep1}, we get:
\begin{equation}
\begin{gathered}
i_r \equiv  \frac{I_r-I_{exc}}{I_c-I_{exc}} = 
\left.
\sqrt{ \frac{\sqrt{1+4g^2}-1}{2g^2}}
    \right|_{\mathrm{model}\,\,\,(1)}
,
\left.
\begin{cases}
    \frac{2}{\sqrt{g^2+4}}& g<2\\
    \frac{1}{\sqrt{g}}& g\geq 2
    \end{cases}
    \right|_{\mathrm{model}\,\,\,(2)}
,
    \\
\tau_r = 
\left.
\frac{\sqrt{1+4g^2}-1}{\sqrt{1+4g^2}+1}
    \right|_{\mathrm{model}\,\,\,(1)}
,
\left.
\begin{cases}
    \frac{2 g^2}{g^2+4}& g<2\\
    1& g\geq 2
    \end{cases}
    \right |_{\mathrm{model}\,\,\,(2)},
\end{gathered}
\label{eq:iranalytic}
\end{equation}
For weak thermal conductance, $g\gg1$ these expression yields the same value of $i_r \approx 1/\sqrt{g}$ resulting in a large hysteresis, whereas for good heat conductance the result is $i_r \approx 1 -O(g^2)$, asymptotically approaching $1$.

We can comment now on the applicability of the approximation $\sqrt{i_0^2-i_J^2(\tau)} \omega_{RSJ}\gg \gamma$ we used above. This approximation may break when $\tau$ becomes too small, such that $|I_J(\tau_r)-I_0|\ll I_J(0)$. However, the hysteresis size in experiments results in $g\sim 1$ for both models such that $1-\tau\sim~1$ and our assumption is well satisfied for values relevant for our experimental observations.

Another useful quantity is $i_{th}$, where $\tau(i_{th})=1$:
\begin{equation}
    i_{th} =\frac{1}{\sqrt{g}},
\end{equation}
which is true for both models since $I_J(1) = 0$.
For $i<i_{th}$ deviations from the linear IV characteristic will appear; note that for $g>2$ in the model (2) there will be no deviations from linear IV down to the retrapping current.

\begin{figure}[h!]
    \centering
\includegraphics[width=0.5\textwidth]{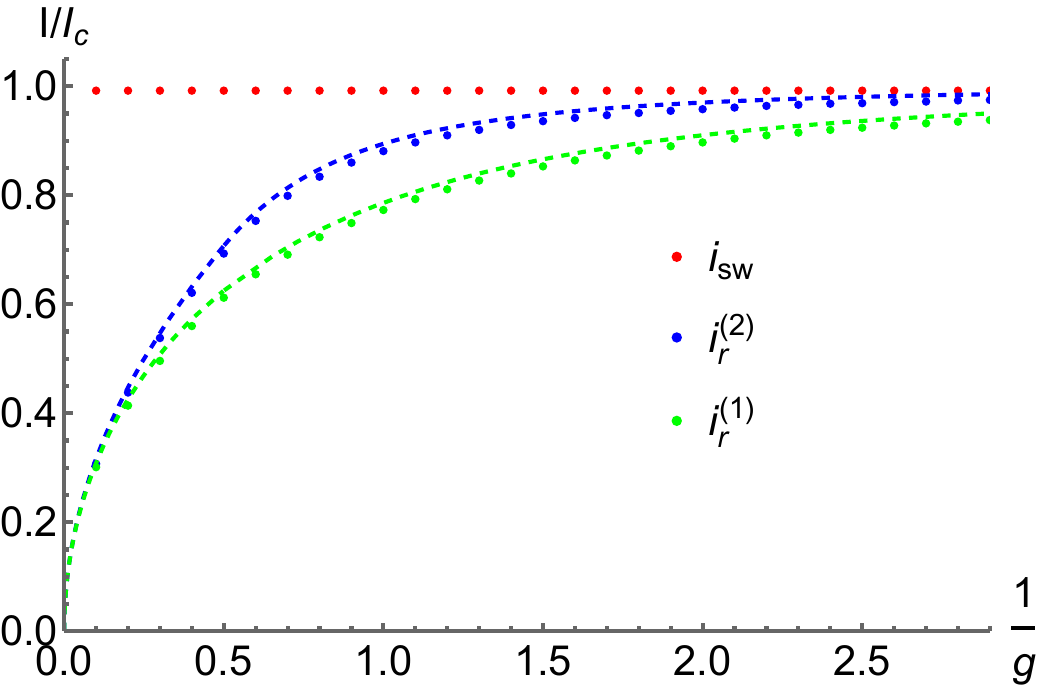}
    \caption{Dependence of switching and retrapping current on $g$, \eqref{eq:gdef}; dashed lines are analytical results \eqref{eq:iranalytic}.
    }
    \label{fig:hyst}
\end{figure}

In Fig.~\ref{fig:hyst} we compare with numerical results obtained from numerical solution of Eqs. (\ref{eq:rsj},\ref{eq:heat}) on increasing/decreasing current. The results are for $\gamma=0.01$ and a varying value of $k$. The external current is sweeped linearly in time from $0$ to $1.3$ on switching or vice versa with rate $1.3/500000$. Results for both models agree well with the analytical results \eqref{eq:iranalytic}. Qualitatively, the dependencies for both models are similar, but quantitative values disagree by about 15 $\%$ at most. 

\subsection{Response to RF}
\label{sec:heat:an_ac}

\subsubsection{Switching}

Assuming $\omega_{RSJ}\gg \omega_L,\gamma,k$ allows one to make a simple argument about switching based on \eqref{eq:rsjnum}. The initial state of the system is $\tau=0, i_J(0) = 1$ and one can consider that $i_{SC}(t)$ changes adiabatically. Finite voltage will appear then when $i_{SC}(t)>1$ at any moment, i.e. $i_{sw}$ is set by $\max_t i_{SC}(t)=1$.
Solving Eq. \eqref{eq:indnum} in the $R_J\ll R$ limit for $i(t) = i_{DC}+i_{rf}\sin(\omega t)$ we obtain:

\begin{equation}
\begin{gathered}
i_{SC}(t) = i_{DC} + i_{rf} \frac{\omega_L}{\sqrt{\omega_L^2+\omega^2}}\sin(\omega_t-\psi_\omega),
\\
\psi_\omega =  \arctan \frac{\omega}{\omega_L}.
\end{gathered}
\label{eq:isc}
\end{equation}
From the criterion $\max_t i_{SC}(t)=1$ derived above, we obtain the switching DC current:
\begin{equation}
\begin{gathered}
i_{sw} = 1- i_{rf} \frac{\omega_L}{\sqrt{\omega_L^2+\omega^2}}.
\end{gathered}
\end{equation}
This result implies a $1/\omega$ asymptotic approach to the DC value at large frequencies $\omega\gg\omega_L$. This results compares well to the numerical results in Fig.~\ref{fig:freq}. We can fit the high-frequency $I_s^-$(above $2.5$ MHz) curves to get $2.7$ MHz for panel (a) and $4.8$ MHz for panel (c), whereas $\omega=_L = 2.5$ MHz and $4.9$ MHz for the two cases.

We note that due to limited frequency range, a scaling analysis of experimental data at large $\omega$, similar to the retrapping one, is not feasible for the switching current.

\subsubsection{Retrapping: General properties}
Here we show below that for any analytic function $i_J(\tau)$ the high-frequency asymptotic behavior of $i_r(\omega)$ should have a wide range of $i_r(\omega)\propto 1/\omega^2$ behavior.
In the following we will use the approximation $i_J(\tau)\ll i_{SC}(t)$ that is well motivated by the experimental results. In particular, the voltage $\langle\dot \varphi\rangle \approx \sqrt{i_{SC}^2-i_J^2}$ on the junction deviates from simple Ohm's law $V = I R_J$ due to nonzero $i_J$. However, in the experiments there are only very small deviations from the linear behavior even extremely close to retrapping, suggesting that $i_J(\tau)\ll i_{SC}(t)$.

Let us simplify the problem for the case where $\gamma,k\ll1$ (Eq. \eqref{eq:num}), corresponding to $\omega_{RSJ}$ being much larger than the other frequency scales. The second (RSJ) equation can then be solved, assuming that RF frequency is also small. As a result, one gets $\langle\dot \varphi^2\rangle \approx i_0 \sqrt{i_0^2-i_J^2}$, such that a closed-form equation for $\tau$ is obtained. In the limit $i_J\ll i_{SC}$ the resulting equation is:
\begin{equation}
    \dot \tau \approx -\gamma  \tau +k i_0^2 -k \frac{i_J^2(\tau)}{2}-k \frac{i_J^4(\tau)}{8 i_{SC}^2(t)}+ ki_{SC}^2(t).
    \label{eq:retr_approx}
\end{equation}

This equation allows to discuss the large-frequency asymptotic of the retrapping current. In particular, we use the ansatz for $\tau(t) = \tau_0(t) + \tau_1(t)$, where $\tau_1(t)$ is an oscillating function with frequency of the order $\omega$ and $\tau_0(t)$ is a slowly (on the scale of $1/\omega$) varying function. Using $\omega\gg k, \gamma$, the equation for $\tau_1(t)$ reads:
\begin{equation}
\dot \tau_1 \approx k [2 i_{DC} \tilde{i}_{rf} \sin(\omega t)+ -\tilde{i}^2_{rf} \cos(2\omega t)/2],
\end{equation}
with the solution:
\begin{equation}
\tau_1 \approx -\frac{2k i_{DC} \tilde{i}_{rf}}{\omega} \cos(\omega t) -\frac{k \tilde{i}_{rf}^2}{4\omega}\sin(2\omega t)
\end{equation}
where we used the form of $i_{SC}(t)$ from \eqref{eq:isc}.

To obtain the equation for $\tau_0$, we average the equation \eqref{eq:retr_approx} over time period $2\pi/\omega$, such that the oscillatory terms average out to zero. Moreover, for large $\omega$, we can expand the equation in $\tau_1$. Assuming $i_J(\tau)$ to be an analytic function, we get:

\begin{equation}
\begin{gathered}
\dot \tau_0
= -\gamma  \tau_0+ki_0^2 -k \frac{i_J^2(\tau_0)}{2}
 +k i_{DC}^2 + \tilde{i}^2_{rf}/2
 \\
 -k (i_J'^2(\tau_0)+i_J''(\tau_0)i_J(\tau_0)+O([i_J/i_{SC}]^2) \frac{\langle\tau_1^2\rangle_t}{2}
 -ki_J^3(\tau_0)i_J'^2(\tau_0)\left\langle\frac{\tau_1(t)}{2 i_{SC}^2(t)}\right\rangle.
 \end{gathered}
\end{equation}
At large $\omega$, the second line can be treated perturbatively in $1/\omega$,
shifting $i_r$ to a larger value (i.e., increase in $i_{DC}$ is required to compensate for it).
Importantly, the first term on the second line scales as $1/\omega^2$, while the second one - as $1/\omega$. The latter is however, accompanied by a small factor $[i_J/i_{SC}]^2$ (and for weak $RF$ drive, a factor of $\tilde{i}_{rf}/i_{DC}\ll1$ in addition to that) such that in practice there is a wide region where $i_r = i_r +O(1/\omega^2)$. 
 
\subsubsection{Retrapping: model analysis}

We now consider the asymptotic behavior of the retrapping current for the two models of $i_J(\tau)$: \eqref{eq:ij_tdep1} and \eqref{eq:ij_tdep2}. Importantly, the outcome will depend qualitatively on the from of $i_J(\tau)$.

{\bf Model 1} \eqref{eq:ij_tdep1}:
\begin{equation}
    \dot \tau = -\gamma  \tau +k i_J(\tau)^2
    \approx 
    -\gamma  \tau - k \frac{(1-\tau)^2}{2}+ ki_0^2(t).
\end{equation}

We can solve this equation perturbatively in RF amplitude $i_0 = i +i_{RF} \sin(\omega t)$ for frequencies much larger than $\gamma$: $\tau \approx \tau_0(t) +\tau_1(t)$, where $\tau_1(t)$ is an oscillating function with frequency of the order $\omega$ and $\tau_0(t)$ is a slowly (on the scale of $1/\omega$) varying function. One gets $\tau_1(t) \approx - \frac{2 k i i_{RF}\cos \omega t}{\omega}$. The equation for the "slow" component $\tau_0$ takes the form:
\begin{equation}
    \dot \tau_0 = -\gamma  \tau_0  - k \frac{(1-\tau_0)^2}{2}+ k i^2 +\frac{k i_{rf}^2}{2}- k\left(\frac{k i i_{RF}}{\omega}\right)^2.
\end{equation}
One notices that for $i$ below a critical value, the r.h.s. is purely negative and thus $\tau$ goes to zero eventually (retrapping occurs). The corresponding current value is:
\begin{equation}
\begin{gathered}
         k i_r^2 +\frac{k i_{rf}^2}{2}- k\left(\frac{k i i_{RF}}{\omega}\right)^2 = \gamma \left(1-\frac{\gamma}{2k}\right),
     \\
i_r  = \sqrt{\frac{\gamma}{k}\left(1-\frac{\gamma}{2k}\right)-\frac{i_{rf}^2}{2}+ \left(\frac{ki i_{RF}}{\omega}\right)^2}\approx
i_{r0}
-\frac{i_{rf}^2}{4}
+
\frac{k^2 i_{r0}^2 i_{RF}^2}{2\omega^2}.
\end{gathered}
\end{equation}
There are two important observations: (1) Retrapping current depends as $1/\omega^2$ at large $\omega$ (2) the coefficient is independent of $\gamma$. In hindsight, this is not unreasonable since we're discussing $\omega\gg \gamma$.
\\
Numerical solutions seem to conform to this prediction; the prefactor differs by a factor 2-3 though. The empirical fitting function that fits numerical data well is:
\begin{equation}
    I_{rf}(\omega) = I_{rf}^{\infty} + \frac{(I_{rf}^{0}- I_{rf}^{\infty})\Gamma^2}{\omega^2+\Gamma^2},
\end{equation}
where $\Gamma^2\sim i_{r0} i_{RF} k^2$. $i_{RF}$ should be equal to $(I_{rf}^{0}- I_{rf}^{\infty})$, so this still allows to determine $\gamma$ using the hysteresis width.
\\
\\
{\bf Model 2}:
\\
\\
We will demonstrate now that for model (2), the asymptotic behavior of $i_r(\omega)$ is $\sim \frac{1}{\omega}$ for $k>2\gamma$.

The solution can be found analytically as follows. The equation is:
\begin{equation}
    \dot \tau \approx 
    -\gamma  \tau - k \frac{1-\tau}{2}\theta(1-\tau)+a+b\sin(\omega t),
\end{equation}
where $a=k i_0^2$, $b=2 k i_0 i_{rf}$ and we neglected the $k i_{rf}^2 \sin(2\omega t)$ term, which is possible in the limit $i_{rf} \ll i_0$. Note also that the response of $\tau$ to this term is also suppressed by a factor of $1/4$ with respect to $\sin(\omega t)$. In the absence of RF drive, the retrapping current value is set by $\gamma = a$ corresponding to $\tau=1$.

We assume that over one period of the RF drive there are two time intervals, where one has $\tau>1$ and the other, $\tau<1$. The general solution is then a periodic piecewise function with period $2\pi/\omega$. We divide a single period in two parts as noted above; for $\omega t\in (\Phi_0-\Phi/2,\Phi_0+\Phi/2)$ one has $\tau<1$, while for $\omega t\in (\Phi_0+\Phi/2,\Phi_0-\Phi/2+2\pi)$ one has $\tau>1$. $\Phi_0$ and $\Phi$ are constants to be determined by the boundary conditions. For a single period  $\omega t\in (\Phi_0-\Phi/2,\Phi_0-\Phi/2+2\pi)$.

\begin{equation}
\tau= \begin{cases}
    C_0 e^ {\frac{k-2 \gamma}{2} [t-\Phi_0]}- \frac{a-k/2}{\delta}
-b \frac{\omega \cos \omega t +\delta\sin \omega t}{\omega^2+\delta^2}
& t \omega \in (\Phi_0 -\Phi/2,\Phi_0 +\Phi/2)\\
    C_1 e^ {-\gamma [t-\Phi_0-\pi]}+ \frac{a}{\gamma}
-b \frac{\omega \cos \omega t  -\gamma \sin \omega t}{\omega^2+\gamma^2}    & t \omega \in (\Phi_0 +\Phi/2,\Phi_0 -\Phi/2+2\pi),
    \end{cases}
\end{equation}
where
\begin{equation}
    \delta = k/2-\gamma,
\end{equation}
and we consider the case $k>2 \gamma$. The boundary conditions $\tau=1$ at the region's edges are
\begin{equation}
\tau([\Phi_0-\Phi/2]/\omega) = \tau([\Phi_0+\Phi/2]/\omega)=\tau([\Phi_0-\Phi/2+2\pi]/\omega)=1.
\end{equation}
This results in the following equations:
\begin{equation}
    \begin{gathered}
        C_0 \sinh \left( \frac{\Phi\delta}{2 \omega}  \right)
        +b \frac{\omega}{\omega^2+\delta^2} \sin \Phi_0 \sin \frac{\Phi}{2}
         -b \frac{\delta}{\omega^2+\delta^2}  \cos \Phi_0 \sin \frac{\Phi}{2} =0,
        \\
        C_0 \cosh \left( \frac{\Phi\delta}{2 \omega}\right)
        +\frac{a-k/2}{\gamma-k/2}
        -b \frac{\omega}{\omega^2+\delta^2} \cos \Phi_0 \cos \frac{\Phi}{2}
         -b \frac{\delta}{\omega^2+\delta^2}  \sin \Phi_0 \cos \frac{\Phi}{2}=0,
         \\
        C_1 \sinh \left( \gamma \frac{2\pi -\Phi}{2\omega}  \right)
        +b \frac{\omega}{\omega^2+\gamma^2} \sin \Phi_0 \sin \frac{\Phi}{2}
         +b \frac{\gamma}{\omega^2+\gamma^2}  \cos \Phi_0 \sin \frac{\Phi}{2}=0,
        \\
        C_1 \cosh \left( \gamma \frac{2\pi -\Phi}{2\omega}  \right)+ \frac{a-\gamma}{\gamma}
        -b \frac{\omega}{\omega^2+\gamma^2} \cos \Phi_0 \cos \frac{\Phi}{2}
         +b \frac{\gamma}{\omega^2+\gamma^2}  \sin \Phi_0 \cos \frac{\Phi}{2}=0.
    \end{gathered}
\end{equation}
The coefficients $C_0,C_1$ can be excluded first resulting in two equations:
\begin{equation}
    \begin{gathered}
        \frac{b \omega \delta}{(\omega^2+\delta^2)(a-\gamma)} 
        \left[\cosh \left( \frac{\Phi\delta}{2 \omega}  \right)
        \sin \frac{\Phi}{2} \left(\sin \Phi_0 - \frac{\delta}{\omega} \cos \Phi_0\right)\right.
        \\
        \left.
       + \sinh \left( \frac{\Phi\delta}{2 \omega}  \right)
        \cos \frac{\Phi}{2} \left(\cos \Phi_0 + \frac{\delta}{\omega} \sin \Phi_0\right)
        \right]+\sinh \left( \frac{\Phi\delta}{2 \omega}  \right)=0,
        \\
        \frac{b \omega \gamma}{(\omega^2+\gamma^2)(a-\gamma)} 
        \left[\cosh \left( \gamma \frac{2\pi -\Phi}{2\omega} \right)
        \sin \frac{\Phi}{2} \left(\sin \Phi_0 + \frac{\gamma}{\omega} \cos \Phi_0\right)\right.
        \\
        \left.
       + \sinh \left( \gamma \frac{2\pi -\Phi}{2\omega} \right)
        \cos \frac{\Phi}{2} \left(\cos \Phi_0 - \frac{\delta}{\omega} \sin \Phi_0\right) 
        \right]-\sinh \left( \gamma \frac{2\pi -\Phi}{2\omega} \right)=0.
    \end{gathered}
\end{equation}
In the limit $\omega\to \infty$ one expects $a_{cr}-\gamma \to 0$, where for $a_{cr} = k i_r^2$, (i.e. the relevant current values are close to the undriven retrapping current set by $a=\gamma$). As a result, for both terms in the equation above to be of the same order in this limit, the content of the brackets has to go to zero, implying that either $\Phi_0\to0$ or $\Phi\to 0$. The second possibility is self-contradictory, as for $\Phi=0$ the system is always at $\tau>1$. Thus, $\Phi_0\to0$ in the $\omega\to \infty$ limit. In the leading order, we assume that $\Phi_0\propto 1/\omega$ and show below that this leads to a consistent solution in the $\omega\to \infty$ limit.

For $\Phi_0\propto 1/\omega$, keeping leading terms in $1/\omega$ in the first equation results in:
\begin{equation}
    \Phi_0 \approx \frac{\delta}{\omega}
    -
    \frac{ \frac{\Phi\delta}{2 \omega}}{\sin \frac{\Phi}{2}}
    \left(\frac{(a-\gamma)\omega}{b\delta} +\cos \frac{\Phi}{2}\right),
\end{equation}
which requires $a-\gamma\propto 1/\omega$ to be shown below. We next get for $\Phi$:
\begin{equation}
    k \sin \frac{\Phi}{2}+ (2\pi \gamma -k \Phi /2)\cos \frac{\Phi}{2} = \frac{2\pi (a-\gamma)\omega}{b}.
\end{equation}
The function on the l.h.s. is bounded from above and below, so the solution exist only for a range of $\Phi$. The extrema of the l.h.s. occur at $\Phi = 0,2\pi,\frac{4\pi \gamma}{k}$, where $\Phi=0$ and $\Phi=2\pi$ correspond to maxima with values $2\pi \gamma$ and $2\pi(k/2-\gamma)$, respectively and $\Phi=\frac{4\pi \gamma}{k}$ is the minimum with the value $k \sin \frac{2\pi \gamma}{k}$. Thus, a nontrivial solution exists for $k \sin \frac{2\pi \gamma}{k}<\frac{2\pi (a-\gamma)\omega}{b}< \max[2\pi \gamma,2\pi(k/2-\gamma)]$. Using the definitions of $a$ and $b$ we find $\frac{2\pi (a-\gamma)\omega}{b} \approx \frac{4 \pi \omega (i_0-\sqrt{\gamma/k})}{i_{rf}}$, which yields
\begin{equation}
    i_r(\omega\to \infty) \approx
    %_{\omega\to \infty} 
    \sqrt{\frac{\gamma}{k}}
    +
    \frac{k i_{rf} \sin \frac{2 \pi \gamma}{k}}{2 \pi\omega}
\end{equation}
This leads to the empirical fitting function
\begin{equation}
\begin{gathered}
        I_{rf}(\omega) = I_{rf}^{\infty} + \frac{(I_{rf}^{0}- I_{rf}^{\infty})\Gamma^{(2)}}{\sqrt{\omega^2+(\Gamma^{(2)})^2}},
    \\
    \Gamma^{(2)} =     \frac{k \sin \frac{2 \pi \gamma}{k}}{2 \pi}
\end{gathered}
    \label{eq:fitirf2}
\end{equation}

We can compare this to numerical results in Sec. \ref{sec:numerics}. For $\gamma=0.0025$ numerical fitting of the results in Fig.~\ref{fig:freq} yields $\Gamma^{(2)}\approx 0.49$ MHz, while the expression \eqref{eq:fitirf2} yields $0.4$ MHz. The discrepancy can be attributed to the influence of the kinetic inductance, so that $\omega_L\gg \gamma, k$ is only approximate. On the other hand, for data in panel (c) of Fig.~\ref{fig:freq} the model fails, yielding $0.025$ MHz, while the characteristic scale in the plot is still of the order MHz. The reason for this discrepancy is that for $\gamma/k>1/2$ the model exhibits a $\sim 1/\omega^2$ at $\omega\gg 1$ behavior as does model (1) or any analytical $I_J(T)$ dependence will.

\subsection{Effects of finite $R$}
\label{sec:finiteR}

We can now also assess the effects of a finite $R_J/R_{bulk}$ ratio in Eq. \eqref{eq:indnum}. As discussed above, for $i>i_J$ $\dot \phi$ evolves with time \eqref{eq:phidotadiab} at high frequencies of the order $\omega_{RSJ}$ for all regimes of interest. Therefore,

Eq. \eqref{eq:indnum} can be solved as:
\begin{equation}
    i_{SC}(t) = \omega_L e^{-\omega_L t} \int_{-\infty}^t dt' \left[i(t') - \dot \varphi \frac{R_J}{R_{bulk}}\right] \approx \tilde{i}(t) - \langle\dot \varphi\rangle \frac{R_J}{R_{bulk}},
\end{equation}
resulting in a renormalization of the current in \eqref{eq:rsjnum} $i_{SC}(t)\to \tilde{i}(t) - \langle\dot \varphi\rangle \frac{R_J}{R_{bulk}}$. We now use the expression for $\langle\dot \varphi\rangle$ from Eq.\eqref{eq:rsjnum}:
\begin{equation}
    \langle\dot \varphi\rangle  = \sqrt{\left(\tilde{i} - \langle\dot \varphi\rangle \frac{R_J}{R_{bulk}}\right)^2-i_J^2}.
\end{equation}
Solving the equation for $\langle\dot \varphi\rangle$ we obtain (only one of two roots is physical):
\begin{equation}
    \langle\dot \varphi\rangle = \frac{\sqrt{\tilde{i}^2 - i_J^2\left(1-\frac{R_J^2}{R_{bulk}^2}\right)} - \tilde{i} \frac{R_J}{R_{bulk}}}{1-\frac{R_J^2}{R_{bulk}^2}}.
    \label{eq:RRj}
\end{equation}
Clearly, a small $R_J$ leads to a weak current renormalization. On the other hand, \eqref{eq:RRj} is well-defined for all values of $R_J$ and for large $R_J$ is approximately equal to $\langle\dot \varphi\rangle\approx (\tilde{i}-i_J)/(R_J/R_{bulk})$, implying that current through the junction is always renormalized to value close to $i_J$, which would strongly suppress hysteretic effects.

\clearpage

\end{document}